\documentclass[a4paper]{article}
\usepackage[left=2.5cm,right=2.5cm,top=2.5cm,bottom=2.5cm]{geometry}
\usepackage{setspace}
\usepackage[english]{babel}
\usepackage[utf8x]{inputenc}
\usepackage[T1]{fontenc}
\usepackage{color}

\usepackage{natbib}
\usepackage[dvipsnames]{xcolor}
\usepackage{graphicx}
\usepackage[export]{adjustbox}
\usepackage{subcaption}
\usepackage{siunitx}
\graphicspath{{Figures/}}
\usepackage{verbatim}

\usepackage{mathtools}
\usepackage{bm}
\usepackage{mathrsfs}
\usepackage{mymathscommands} 
\numberwithin{equation}{section}

\newcommand{\EcrossB}{$\bm{E}\times \bm{B}$ }

\usepackage{authblk}

\title{Stability of scrape-off layer plasma: a modified Rayleigh-B{\'e}nard problem }
\author[1]{F. Wilczynski}
\affil[1]{EPSRC CDT in Fluid Dynamics, University of Leeds, Leeds LS2 9JT, United Kingdom} 
\author[2]{D. W. Hughes}%
\affil[2]{School of Mathematics, University of Leeds, Leeds LS2 9JT, United Kingdom}
\author[3]{S. Van Loo}
\affil[3]{School of Physics and Astronomy, University of Leeds, Leeds LS2 9JT, United Kingdom}
\author[4]{W. Arter}
\author[4]{F. Militello}
\affil[4]{CCFE, Culham Science Centre, Abingdon OX14 3DB, United Kingdom}
\date{}

\begin{document}
\maketitle

\begin{abstract}
We present a linear stability analysis of a two-dimensional fluid model used to study the plasma dynamics in the scrape-off layer of tokamaks. The model equations are based on the Braginskii fluid equations under the assumptions of drift ordering and an electrostatic plasma. The model also employs the common slab geometry approximation, whereby the magnetic field is assumed constant and straight, with the effects of curvature reintroduced as effective gravitational terms. We demonstrate that the governing plasma equations for the scrape-off layer can be viewed as describing a thermal convection problem with additional effects. The new features include a non-uniform basic state gradient, linear damping terms, and additional advective terms. We characterise the conditions at the onset of instability, and perform an extensive parameter scan to describe how the stability threshold varies as a function of plasma parameters. 

\end{abstract}

\section{Introduction}
In magnetic confinement devices, boundary turbulence is responsible for transporting plasma and energy from the well-confined region towards the material surfaces. It has been universally observed that the boundary plasma is characterised by large, intermittent fluctuations, often called filaments or blobs, which dominate the particle transport and enhance the plasma interaction with the surrounding material boundaries \citep{krasheninnikov2008recent, dippolito2011review}. This is problematic as plasma-wall interaction can potentially enhance erosion and shorten the lifetime of the machine. A full understanding of filament dynamics is therefore essential for the successful operation of future fusion experiments and reactors.

Over the last two decades, significant experimental and theoretical work has been devoted to understanding the fundamental mechanisms governing the dynamics in the scrape-off layer of magnetic fusion devices. Various numerical models have been developed and used to study boundary turbulence and filament dynamics, both in 2D and 3D; some of the notable examples include STORM \citep{easy2014three}, HESEL \citep{nielsen2015simulation}, GBS \citep{ricci2012simulation} and TOKAM3X \citep{tamain2010tokam}. These models are derived from the Braginskii fluid equations, assuming drift ordering and electrostatic plasma. At the very basic level, the model equations consist of evolution equations for density conservation, plasma vorticity (which determines the plasma potential), as well as parallel ion and electron momenta. For practical purposes, several further approximations are introduced to simplify the equations: for example, a common simplification is the slab geometry approximation, in which the magnetic field is assumed constant and straight, with the effects of curvature reintroduced through artificial terms. Although, in general, these approximations vary from model to model (for an in-depth discussion of key differences between the models, see \cite{militello2016multi, riva2016blob}), their essential features remain the same, and good agreement has been found between the different models \citep{militello2016multi}. While 3D codes have emerged as the new standard in the past few years, 2D codes are still commonly employed, as they have the advantage of greatly simplifying the analysis of cross-field SOL transport, while still retaining the fundamental properties of the underlying physics. Such 2D models, constructed by invoking \textit{ad hoc} closures for the dynamics in the parallel direction, were shown to be able to capture several experimentally measured features of the midplane SOL plasma  \citep{garcia2004computations, myra2008transport, russell2009saturation, fundamenski2007dissipative, militello2013experimental, bisai2005edge}.

In this paper we revisit the known analogy between the instability of SOL plasma and that of Rayleigh-B\'enard convection (RBC) \citep{berning2000bifurcations, ghendrih2003theoretical, garcia2006two}, thereby demonstrating that this analogy is not as clear-cut as previous literature has suggested. Previous considerations have restricted attention to the paradigmatic model for two-dimensional thermal convection; i.e.\ it is assumed from the outset that the plasma edge can be modelled using the conventional Rayleigh-B{\'e}nard equations (sometimes augmented by the inclusion of heuristic dissipation terms to account for the presence of particle sinks at the sheath of the SOL \citep{bian2003blobs, garcia2005mechanism, garcia2006radial, aydemir2005convective}). As a consequence, a number of features that are relevant to the magnetized plasma problem are neglected. Here, we shall begin with the two-dimensional SOL equations and show that these can be `naturally' reduced to a modified convection problem. An intuitive way to explore the analogy between these two systems is through the means of linear stability analysis; the linear stability properties of RBC are well understood, and it is therefore natural to ask how these stability properties change in the presence of the additional plasma-related features. In addition, the linear stability analysis is a valuable first step in studying complicated fluid systems, providing potentially important pointers to the nonlinear regime.

We consider a well-established two-dimensional fluid model for SOL plasma, described in \citet{easy2014three, easy2016investigation}. It is a two-field (density, vorticity) model describing the plasma dynamics perpendicular to the magnetic field, which invokes the sheath dissipation assumption in order to provide closure for the current along field lines. Although such models are fairly standard in SOL turbulence, there has been surprisingly little work exploring thoroughly their linear stability properties. Furthermore, previous linear stability calculations have limited attention to perturbations that are periodic in both radial ($x$) and poloidal ($y$) directions, expressing perturbations as simple Fourier modes \citep{vilela2017analytical}, or have neglected the radial variation completely \citep{bisai2004simulation}; such treatments do not take into account the influence of boundary conditions on the stability properties. In hydrodynamics, it is well known that the choice of boundary conditions can affect both the stability properties of the system and the nature of the solutions. It is therefore of interest to begin classifying these effects with regard to the plasma problem. In our analysis, periodicity is assumed only in the poloidal direction, while the radial extent is bounded; the radial structure of the perturbation is then determined as the solution of an eigenvalue problem. We will find that, owing to the appearance of the explicit $x$ dependence of the coefficients in the problem, the radial structure of solutions can become highly localised --- a behaviour that cannot be recovered when the radial direction is treated as periodic. The emphasis of this study is to characterise the conditions at the onset of instability. We perform an extensive parameter scan to describe how the stability threshold varies as a function of plasma parameters. 

In addition to solving the linear eigenvalue problem numerically, we will use the analogy to RBC to construct a reduced linear system that allows an analytical solution, and we will compare those against the solutions to the full problem. We will find that the reduced system provides useful insight into the qualitative behaviour of the full problem; in particular, it accurately predicts the changes to the stability threshold subject to variations of plasma parameters. Furthermore, we will identify an approximate range in the parameter space for which there is good quantitative agreement between the full and reduced systems.

The remainder of the paper is structured as follows. In section \ref{sec:gov_eqns} we introduce the governing equations and describe the relationship with the Rayleigh-B{\'e}nard problem. For completeness we include a description of the three-dimensional version of the model and briefly demonstrate the procedure of reducing the equations to two dimensions. Section \ref{sec:LSA} contains the formulation of the linear eigenvalue problem, and the results are analysed in section \ref{sec:results}. Finally, the conclusions of the paper are summarized in section \ref{sec:discussion}.

\section{Mathematical formulation} \label{sec:gov_eqns}
\subsection{Governing equations} 
As discussed in the Introduction, we consider here the electrostatic drift fluid model of \cite{easy2014three, easy2016investigation}. The geometry is simplified to a local slab with a uniform magnetic field $\bm{B} = B\bm{\hat{z}}$; the effects of magnetic curvature and inhomogeneity of $\bm{B}$ are then represented through additional effective gravity terms acting in the radial direction. Coordinates $x$ and $y$ represent respectively the effective radial and poloidal directions. The system is governed by evolution equations for the plasma density $n$, vorticity $\omega = \nabla_\perp^2 \varphi/B$, where $\varphi$ is the electrostatic potential, parallel ion velocity $v_{\parallel i}$ and parallel electron velocity $v_{\parallel e}$:
\begin{gather}
\frac{m_i n}{B} \brac{\pdev{}{t}+\mb{v}_E \cdot \nabla  + v_{\parallel i}\nabla_\parallel } \omega=   
\nabla_\parallel j_\parallel 
- \frac{e g}{\Omega_i} \pdev{n}{y}
+  \frac{ n m_i  \nu_i }{B} \nabla_\perp^2 \omega , \label{eqn:vorticity3d} \\
\pdev{n}{t} + 
\mb{v}_E \cdot \nabla n  + \nabla_\parallel \brac{{v}_{\parallel e} n }  = \frac{g n}{B c_s^2} \pdev{\varphi}{y} - \frac{g}{\Omega_i} \pdev{n}{y}
+ D \nabla_\perp^2 n
+s_n, 
\label{eqn:density3d}
\\
m_e n \brac{\pdev{}{t} + \mb{v}_E \cdot \nabla + v_{\parallel e}\nabla_{\parallel} } v_{\parallel e} = en \nabla_{\parallel} \varphi - T_e \nabla_{\parallel} n + e n \eta j_\parallel - m_e s_n v_{\parallel e} ,
\label{eqn:parallel_electron}
\\
m_i n \brac{\pdev{}{t} + \mb{v}_E \cdot \nabla + v_{\parallel i}\nabla_{\parallel} } v_{\parallel i} = -en\nabla_{\parallel} \varphi - en \eta j_\parallel - m_i s_n v_{\parallel i} . \label{eqn:parallel_ion}
\end{gather}
Here, $\mb{v}_E = B^{-1}\brac{\bm{\hat{b}} \times \nabla\varphi}$ is the $\bm{E} \times \bm{B}$ drift velocity, and $j_\parallel = en\brac{v_{\parallel i} - v_{\parallel e}}$ is the parallel current density; $s_n$ is a particle source, $\nu_i$ represents the effective cross field kinematic viscosity of ions, $D$ is the particle  diffusion coefficient, $e$ is the elementary unit charge, $m_i$ is the ion mass, $c_s = \sqrt{T_e/m _i}$ is the sound speed, and $T_e$ is the electron temperature in Joules. The ion gyrofrequency $\Omega_i = eB/m_i$ is related to the sound speed through the gyroradius $\rho_s = c_s/\Omega_i$. The parameter $g = 2c_s^2/R_c$ represents the effective gravitational acceleration that captures the influence of magnetic gradients and curvature, and $R_c$ is the radius of curvature (typically the major radius of the machine). The two curvature terms in the density equation \eqref{eqn:density3d} represent compressibility of the $\bm{E} \times \bm{B}$ drift, and compressibility of diamagnetic drift respectively. In the vorticity equation \eqref{eqn:vorticity3d} the curvature term represents the divergence of the diamagnetic current. Note that under the cold ion assumption the diamagnetic current is composed entirely of the electron diamagnetic drift.

The governing equations can be simplified to a two-dimensional system by implementing a suitable closure for the current along the field lines. Two commonly employed closures are what are known as the sheath dissipation closure and the vorticity advection closure. The majority of studies invoke the sheath-limited model as it has been demonstrated to perform better than the vorticity advection closure at capturing the plasma dynamics associated with blobs \citep{easy2014three, riva2016blob}. The sheath dissipation closure assumes negligible gradients of density and potential in the parallel direction and also that parallel current is regulated by the following sheath boundary conditions \citep{omotani2015effects, yu2006two}:
\begin{gather}
v_{\parallel e} (z = \pm l_\parallel) = \pm   c_s \exp \brac{-\frac{e}{T_e} \varphi } ,
\\
 v_{\parallel i} (z = \pm l_\parallel) = \pm   c_s ,
\\
j_\parallel(z = \pm l_\parallel) = \pm  e c_s n_e \brac{1 - \exp \brac{-\frac{e}{T_e} \varphi } } ,
\end{gather}
where $l_\parallel$ is the parallel SOL connection length (typically the mid-plane to target distance). To obtain the two-dimensional set of equations we integrate equations \eqref{eqn:vorticity3d} and \eqref{eqn:density3d} along field lines between $z = -l_\parallel$ and $z=+l_\parallel$, and apply the sheath dissipation closure; we assume that density and electrostatic potential are uniform along the $z$--direction. Equations \eqref{eqn:vorticity3d} and \eqref{eqn:density3d} then give
\begin{gather}
\brac{\pdev{}{t}+\mb{v}_E \cdot \nabla   } \omega=   
\frac{1}{l_\parallel} c_s \Omega_i \brac{1 - \exp \brac{-\frac{e}{T_e} \varphi } }
- \frac{ g}{n} \pdev{n}{y}
+   \nu_i  \nabla^2 \omega, \label{eqn:vorticity2d}
\\
\brac{\pdev{}{t} + 
\mb{v}_E \cdot \nabla  } n =
-\frac{1}{l_\parallel} n c_s \exp \brac{-\frac{e}{T_e} \varphi }
+ \frac{g n}{B c_s^2} \pdev{\varphi}{y} - \frac{g}{\Omega_i} \pdev{n}{y}
+ D \nabla^2 n 
+ \frac{c_s}{l_\parallel} N(x) . \label{eqn:density2d}
\end{gather}
Here $N(x)$ is the background density and we have assumed that parallel losses are compensated by the source term \citep{easy2014three, vilela2017analytical}, i.e.\ 
\begin{equation}
\frac{1}{2 l_\parallel} \int_{-l_\parallel}^{+l_\parallel} s_n \intd{z} = \frac{c_s}{l_\parallel} N(x).
\end{equation}
Under the sheath dissipation closure, the evolution is governed completely by equations \eqref{eqn:vorticity2d} and \eqref{eqn:density2d}; equations \eqref{eqn:parallel_electron} and \eqref{eqn:parallel_ion} governing parallel dynamics are no longer relevant. 
The change of variable $\theta = \log \brac{n/n_0}$ allows us to recast equations \eqref{eqn:vorticity2d} and \eqref{eqn:density2d} as
\begin{gather}
\brac{\pdev{}{t}+\mb{v}_E \cdot \nabla } \omega=   
\frac{1}{l_\parallel} c_s \Omega_i \brac{1 - \exp \brac{-\frac{e}{T_e} \varphi } }
- g \pdev{\theta}{y}
+   \nu_i  \nabla^2 \omega, \label{eqn:log_vorticity} 
\\
\brac{ \pdev{}{t} + 
\mb{v}_E \cdot \nabla} \theta  =
-\frac{c_s}{l_\parallel} \exp \brac{-\frac{e}{T_e} \varphi }
+\frac{g}{B c_s^2} \pdev{\varphi}{y} - \frac{g}{\Omega_i} \pdev{\theta}{y}
+ D (\nabla^2 \theta +\modulus{\nabla \theta}^2)  
+\frac{c_s}{l_\parallel} \exp\brac{\Theta(x) - \theta},
\label{eqn:log_density}
\end{gather}
where $\Theta (x) =\log(N(x)/n_0)$. 
Note that the diffusion related term $D\modulus{\nabla \theta}^2$ in \eqref{eqn:log_density} comes from the usual $D \nabla^2 n$ term in the density continuity equation \eqref{eqn:density2d}, which transforms according to $\nabla^2 n / n = \nabla^2 \theta + \modulus{\nabla \theta}^2$ upon the change of variable.
At this point, subject to Bohm normalisation, equations \eqref{eqn:log_vorticity} and \eqref{eqn:log_density} are identical to equations (3a), (3b) of \cite{vilela2017analytical}, although these authors left the source term (the last term on the right hand side of \eqref{eqn:log_density}) unspecified. 

We consider a layer of plasma bounded radially between $x=0$ and $x = h$, where $h$ represents the width of the scrape-off layer. The density $n$ is fixed to $n_0 + \Delta n$ at the inner boundary, and $n_0$ at the outer boundary. We consider a steady basic state with plasma at rest, and assume that the basic state plasma density varies as a function only of the radial coordinate. We describe the basic state by upper case variables; thus $\Phi = 0$ and $n = N(x)$. The vorticity equation \eqref{eqn:log_vorticity} is trivially satisfied while the log density equation \eqref{eqn:log_density} reduces to 
\begin{equation}
\devt{\Theta}{x} + \brac{\dev{\Theta}{x}}^2= 0.
\end{equation}
The basic state log density can thus be expressed as
\begin{equation}
\Theta(x) = \ln\brac{ 1 + \frac{\Delta n}{n_0} \brac{1-\frac{x}{h}} }.
\end{equation}
We now consider small perturbations to this basic state, expressing the potential, vorticity and density in the perturbed state by $\varphi$, $\omega$ and $\Theta +\theta$ respectively. On substituting these expressions into equations~\eqref{eqn:log_vorticity} and \eqref{eqn:log_density} and retaining only the lowest order terms in the perturbations, the linearised form of the equations of motion become
\begin{gather}
\pdev{\omega}{t} = \frac{1}{l_\parallel} c_s \Omega_i \brac{\frac{e}{T_e}\varphi} - {g} \pdev{\theta}{y} + \nu_i \nabla ^2 \omega, 
\label{eqn:vorticity:linear}
\\
\pdev{\theta}{t} - \frac{1}{B}\pdev{\varphi}{y} \dev{\Theta}{x} = 
\frac{g}{B c_s^2}  \pdev{\varphi}{y} - {\frac{g}{\Omega_i} \pdev{\theta}{y}} + D \brac{\nabla^2 \theta+ 2\pdev{\theta}{x} \dev{ \Theta}{x} } 
-\frac{c_s}{l_\parallel} \theta
+\frac{c_s}{l_\parallel} \frac{e}{T_e}\varphi.
\label{eqn:density:linear}
\end{gather} 
Following an approach commonly used in two dimensional simulations \citep{ bian2003blobs, easy2014three,garcia2005turbulence, russell2009saturation}, we assume that perturbation quantities vanish on radial boundaries, i.e.\ 
\begin{equation}
\varphi = \omega = \theta =0 \qquad \text{on} \qquad x = 0,\, h.
\label{eqn:boundary_conditions}
\end{equation}
Periodic boundary conditions are invoked for the poloidal direction.

We now express the governing equations in dimensionless form. Non-dimensionalising $x$ and $y$ by $h$, $t$ by the diffusive time scale ${h^2}/{D}$, $n$ (and $\Delta n$) by $n_0$, and $\varphi$ by $B D$, equations~\eqref{eqn:vorticity:linear} and \eqref{eqn:density:linear} become
\begin{gather}
\pdev{{\omega}}{{t}}  =  
-  {Ra^* \, Pr} \pdev{{\theta}}{{y}} +  Pr {\nabla} ^2 {\omega}
+ \frac{L_\perp^2 \Omega}{L_\parallel}  \varphi ,
\label{vorticity:linear:dimensionless}
\\
\pdev{{\theta}}{t}= 
\brac{\dev{\Theta}{x} + \frac{2 h}{ R_c} }   \pdev{{\varphi}}{{y}}
-  \frac{Ra^* \, Pr}{ \, \Omega} \pdev{{\theta}}{{y}} +  {\nabla}^2 {\theta} + 2\dev{\Theta}{x} \pdev{\theta}{x} 
- \frac{\Omega}{L_\parallel} \theta
+ \frac{L_\perp^2}{L_\parallel} \, \varphi ,
\label{density:linear:dimensionless}
\end{gather}
where $\omega = \nabla^2 \varphi$,
\begin{equation}
\dev{\Theta}{x} = \frac{-\Delta n}{1 +\Delta n (1 - x)},
\label{eqn:basic_state_gradient}
\end{equation}
and where we have introduced the parameters
\begin{align}
Ra^*= \frac{gh^3}{D \nu_i}, \qquad \qquad
Pr = \frac{\nu_i}{D}, \qquad \qquad
\Omega = \frac{\Omega_i h^2}{D}, \qquad \qquad
L_\parallel = l_\parallel/\rho_s, \qquad \qquad
L_\perp = h/\rho_s .
\end{align}
The $Ra^*$ parameter measures the ratio of the strength of the curvature-induced gravitational force to viscous forces. It is similar to the Rayleigh number associated with buoyancy-driven flow, although this analogy is not complete since $Ra^*$ is missing a factor describing the density difference (or temperature difference in convection) across the layer. $Pr$ can be thought of as equivalent to the Prandtl number in the convection problem, but instead of describing the ratio of fluid viscosity to thermal diffusivity, here it represents the ratio of the ion viscosity to the particle diffusivity. $\Omega$ is the gyrofrequency divided by the time scale of diffusion. $\Lpar$ is the normalised measure of parallel connection length, and $\Lperp$ is the normalised measure of the width of the layer.

The physical meaning of the terms in equations~\eqref{vorticity:linear:dimensionless} and \eqref{density:linear:dimensionless} clearly remains unchanged by this scaling. From left to right in the vorticity equation \eqref{vorticity:linear:dimensionless}, the individual terms are linearised versions of the ion polarisation current, the diamagnetic current, the current due to viscosity, and the parallel current to the sheath. 
In the density equation \eqref{density:linear:dimensionless}, the first term on the right hand side represents the density flux due to radial \EcrossB drift velocity, with its two components corresponding to the advection of the background density distribution and the compressibility of the \EcrossB drift. The second term is the density flux due to the diamagnetic drift; the third and fourth terms come from the particle diffusion term in \eqref{eqn:density2d}; the last two terms are representative of parallel losses to the sheath. 

\subsection{Relation to Rayleigh-B{\'e}nard convection}
The fundamental mechanism of interchange drive in boundary plasma has been compared to buoyancy drive in neutral fluids, with reference to Rayleigh-B\'enard convection in particular \citep{berning2000bifurcations, ghendrih2003theoretical, garcia2006two}. Indeed, in their dimensionless form, equations~\eqref{vorticity:linear:dimensionless} and \eqref{density:linear:dimensionless} may be viewed as the equations governing the linear stability of two-dimensional Rayleigh-B{\'e}nard convection (RBC) \citep[e.g.][]{chandrasekhar2013hydrodynamic}, but with the addition of extra terms. By this analogy, the plasma electrostatic potential and plasma vorticity correspond to the fluid streamfunction and fluid vorticity respectively, and the logarithm of plasma density corresponds to fluid temperature. Furthermore, the boundary conditions \eqref{eqn:boundary_conditions} are formally identical to stress-free, fixed temperature boundary conditions in the classical convection problem. These boundary conditions are particularly convenient in the case of the convection problem as they allow an explicit solution and detailed stability analysis. 

The analogous linear convection problem that matches the boundary conditions of the original problem is governed by the equations
\begin{gather}
\pdev{\nabla^2 \psi}{{t}}  =  
-  {Ra^* Pr} \pdev{{\theta}}{{y}} +  Pr \nabla^2 \nabla^2 \psi,
\label{eqn:RB:vorticity}
\\
\pdev{\theta}{t}= 
-\ln(1+\Delta n)   \pdev{\psi}{y}
+  {\nabla}^2 {\theta} ,
\label{eqn:RB:density}
\end{gather}
where $\psi$ is the streamfuction, related to the velocity via $\mb{u} = (0, \pdevi{\psi}{z}, -\pdevi{\psi}{y})$, and $\theta$ is the temperature perturbation. Equations \eqref{eqn:RB:vorticity}, \eqref{eqn:RB:density} govern two--dimensional motion in a plane layer; by convention the vertical direction is identified with the $z$ coordinate. We thus identify the radial ($x$) direction in the plasma problem with the vertical ($z$) direction in the convection problem. Hence the poloidal direction in the plasma problem corresponds to the horizontal direction in the convection problem.

This analogy between the two sets of equations, along with compatible boundary conditions, motivates viewing the system \eqref{vorticity:linear:dimensionless}, \eqref{density:linear:dimensionless} as a modified convection problem, where the modifications can be categorized as follows. First, in RBC the basic state temperature gradient is uniform across the layer, $-\ln(1+\Delta n)$. 
In this case, we could rescale $\theta$ further to write \eqref{eqn:RB:vorticity} and \eqref{eqn:RB:density} in the standard form of RBC, namely
\begin{gather}
\pdev{\nabla^2 \psi}{{t}}  =  
-  {Ra Pr} \pdev{{\theta}}{{y}} +  Pr {\nabla} ^2 \nabla^2 \psi,
\label{eqn:RB:vorticity:standardform}
\\
\pdev{{\theta}}{t}= 
- \pdev{{\psi}}{{y}}
+  {\nabla}^2 {\theta} .
\label{eqn:RB:density:standardform}
\end{gather}
Note now that the Rayleigh number of convection is $Ra = Ra^* \log(1+\Delta n)$. The first term on the right hand side of \eqref{eqn:RB:density} represents the vertical advection of the uniform background temperature gradient. The analogous term in the plasma problem \eqref{density:linear:dimensionless} is composed of two components; the first is the advection of the non--uniform basic state density gradient, while the other (which is representative of the effect of compressibility of the \EcrossB drift) can be thought of as advection of a stabilising uniform gradient. The second modification is that the plasma system \eqref{vorticity:linear:dimensionless}, \eqref{density:linear:dimensionless} includes linear damping terms, proportional to $1/\Lpar$, which are physically representative of particle losses in the parallel direction. Finally, viewed as a modified temperature of RBC, equation \eqref{density:linear:dimensionless} contains two additional advective terms. One corresponds to the diamagnetic drift term which acts to transport $\theta$ perturbations in the poloidal direction; the other can be interpreted as advection of $\theta$ by a spatially dependent flow that is proportional to the basic state density gradient. 

As mentioned above, for the particular choice of boundary conditions, the convection problem can be solved exactly. In contrast, the presence of non-constant coefficients in the plasma problem make it particularly difficult to solve analytically. In the following section, we shall however construct a reduced system that can be solved in the same way as the RBC problem, and compare its solution to the numerical solution of the full system.

\section{Linear stability analysis} \label{sec:LSA}
\subsection{Eigenvalue problem} \label{sec:eigenvalue_problem}
We postulate normal mode solutions to equations~\eqref{vorticity:linear:dimensionless} and \eqref{density:linear:dimensionless} of the form
\begin{align}
\varphi (x,y, t) = \hat{\varphi}(x) \exp \brac{iky + \sigma t} + c.c., \qquad \qquad
\theta (x,y, t) = \hat{\theta}(x) \exp \brac{iky + \sigma t} + c.c.,
\label{eqn:normal_modes}
\end{align}
where $\hat{\varphi}(x)$, $\hat{\theta}(x)$ are complex eigenfunctions, $k$ is the horizontal wave number of a particular normal mode and $\sigma$ is the complex eigenvalue that determines the stability of the system. 
Substituting perturbations \eqref{eqn:normal_modes} into equations \eqref{vorticity:linear:dimensionless} and \eqref{density:linear:dimensionless} results in the linear eigenvalue problem,
\begin{equation}
\brac{\mathscr{L}_1 - \sigma \mathscr{L}_2 }\bm{S} = 0
\label{eqn:eigenvalue_system}
\end{equation}
for the solution vector $\bm{S} = [\hat{\varphi}, \hat{\theta}]^{T}$, where ${T}$ denotes the transpose. The two linear operators $\mathscr{L}_1$ and $\mathscr{L}_2$ are defined by
\begin{gather}
\mathscr{L}_1 =
\begin{bmatrix}
Pr \brac{\ddevt{}{x} -k^2}^2 + \dfrac{L_\perp^2 \Omega}{L_\parallel} & -ik \, Ra^* Pr  \\ 
ik \brac{\ddev{\Theta}{x} + \dfrac{2h}{R_c}}  + \dfrac{L_\perp^2 }{L_\parallel} & -  \dfrac{Ra^* \, Pr}{ \, \Omega} ik +   \brac{\ddevt{}{x} -k^2}  + 2\ddev{\Theta}{x} \ddev{}{x} 
- \dfrac{\Omega}{L_\parallel}
\end{bmatrix},
\label{eqn:linearOperator1}
\\
\mathscr{L}_2 =
\begin{bmatrix}
\brac{\ddevt{}{x} -k^2}  & 0 \\
0 & 1
\end{bmatrix},
\label{eqn:linearOperator2}
\end{gather}
and the boundary conditions for $\hat{\varphi}$ and $\hat{\theta}$ are given by
\begin{equation}
\hat{\varphi} = \devt{\hat{\varphi}}{x} = \hat{\theta} = 0 \qquad \text{at} \qquad x = 0, \, 1.
\label{eqn:boundary_conditions2}
\end{equation}
The eigenvalue problem \eqref{eqn:eigenvalue_system} must be solved numerically; we employ MATLAB's \texttt{bvp4c} routine for solving  boundary value problems. We address the problem of marginal stability; in particular, we seek the minimum value of the density difference $\Delta n$, and the accompanying critical wavenumber $k_c$, for which $Re(\sigma) = 0$. Note that eqns. \eqref{eqn:eigenvalue_system}--\eqref{eqn:boundary_conditions2} shall be referred to as the \textit{full} problem, to be distinguished from the \textit{reduced} problem, which we shall introduce in section~\ref{sec:reduced_system}. Immediately below, we discuss the range of dimensionless parameter values used for these numerical investigations.

\subsubsection{Parameters}
In general, the dimensional parameters in the plasma edge vary from one discharge to another. Thus, rather than stating precise values of the physical parameters, we shall concern ourselves with representative, order-of-magnitude estimates; the following estimates are broadly relevant for the L-mode scrape-off layer in a medium size tokamak. We take estimates for the magnetic field $ B \approx  \SI{1}{T} $, and the radius of curvature  $ R_c  \approx  \SI{1}{m} $ (Ref. \citep{easy2014three}). The width of the SOL, $h$, is typically estimated to be several centimetres (certainly not greater than $\SI{0.1}{m}$), and the parallel connection length  $ l_\parallel  \approx  \SI{10}{m} $. For a typical discharge the edge values for temperature and density are found experimentally to be $T_e  \approx \SI{10}{eV}$, $T_i  \approx  \SI{20}{eV}$, $n_e \approx 10^{18} \,\SI{}{m^{-3}}$ (Ref. \citep{militello2016multi}). Using appropriate formulae (outlined in section \ref{sec:gov_eqns}), these give reasonable estimates for the sound speed $c_s$, the effective gravitational acceleration $g$, the gyro-radius $\rho_s$, and the gyro-frequency $\Omega_i$. The matter of estimating appropriate values of conduction and viscous coefficients on the other hand is far more ambiguous. Depending on the choice of classical \citep{braginskii1965transport} or anomalous values (implied by empirical scaling laws \citep{goldston1984energy}), the particle diffusivity $D$ can range between $\mathcal{O}(10^{-3})$ and $\mathcal{O}(1)$, while the ion viscosity $\nu_i$ can range between $\mathcal{O}(10^{-4})$ and $\mathcal{O}(1)$. Hence it seems that physics uncertainty alone implies that $\Pran$ can range from $10^{-4}$ to $1$, and $\Ra^*$ from $10^5$ to $10^{12}$; similarly, $\Omega$ can range between $10^5$ and $10^8$. This uncertainty in the values of the dimensionless parameters is summarised in Table~\ref{table:parameters}.
\begin{table}
\centering
\begin{tabular}{lc}
\hline
$\Ra^*$ & $\mathcal{O}(10^5)$ -- $\mathcal{O}(10^{12})$  \\
$\Pran $ & $\mathcal{O}(10^{-4})$ -- $\mathcal{O}(1)$ \\
$\Omega $ & $\mathcal{O}(10^5)$ -- $\mathcal{O}(10^8)$ \\
$\Lperp $ & $\mathcal{O}(10)$ \\
$\Lpar $ & $\mathcal{O}(10^3)$  \\
\hline
\end{tabular}
\caption{Range of dimensionless parameters.}
\label{table:parameters}
\end{table}
We focus on the effect of varying $\Ra^*$ and $\Pran$, as the uncertainty in these is greatest, and fix $\Omega = 10^5$, $\Lpar = 5500$, $\Lperp = 55$, $2h/R_c = 0.04$.

\subsection{Reduced linear system}\label{sec:reduced_system}

In the analogous convection problem \eqref{eqn:RB:vorticity:standardform}, \eqref{eqn:RB:density:standardform}, with stress-free, fixed temperature boundary conditions (equivalent to \eqref{eqn:boundary_conditions}), the solutions take the simple sinusoidal form ${\psi}, {\theta} \sim \sin (m \pi z) \exp(\sigma t + iky)$, where $m$ is an integer \citep[cf.][]{chandrasekhar2013hydrodynamic}. The condition for marginal stability is then given by
\begin{equation}
\Ra = \frac{\left( m^2 \pi^2 +k^2 \right)^3}{k^2}.
\label{eqn:RB:marginal}
\end{equation}
Owing to the explicit $x$ dependence of the coefficients in the plasma problem, governed by \eqref{vorticity:linear:dimensionless} and \eqref{density:linear:dimensionless}, simple Fourier modes can no longer be adopted. Hence, to make progress analytically, we construct a reduced linear problem to \eqref{vorticity:linear:dimensionless}, \eqref{density:linear:dimensionless} by extending the Rayleigh--B{\'e}nard problem as far as we can whilst retaining the simplicity of its solutions. To this end, we replace the non-uniform basic state gradient in the first term on the right hand side of \eqref{density:linear:dimensionless} by $-\log(1+ \Delta n)$, and we omit the term $2\Theta' \pdevi{\theta}{x}$ completely. The resulting reduced system is
\begin{gather}
\pdev{{\omega}}{{t}} =  
-  { {\Ra}^*  \Pran} \pdev{{\theta}}{{y}} +  \Pran {\nabla} ^2 {\omega}  + {\frac{\Lperp^2 \Omega}{\Lpar} \varphi}, \label{eqn:reduced:vorticity} \\
\pdev{{\theta}}{t} + \frac{\Ra^* \Pran}{\Omega} \pdev{\theta}{y}
= -\brac{\ln (1+\Delta n) - \frac{2h}{R_c}}  \pdev{\varphi}{y} + {\nabla}^2 {\theta} + {\frac{\Lperp^2}{\Lpar} \varphi} - {\frac{\Omega}{\Lpar} \theta}.
\label{eqn:reduced:density}
k\end{gather}
Note that such a system would arise naturally if we neglect the diffusion related term $D \modulus{\nabla \theta}^2$ in equation \eqref{eqn:log_density}.
The basic state log density would then be linear, given by the solution of $\Theta'' = 0$, and the basic state gradient would be spatially uniform with $\Theta' = -\ln (1 +\Delta n)$. The dimensionless perturbation equations in such a case would then be precisely \eqref{eqn:reduced:vorticity} and \eqref{eqn:reduced:density}.
In contrast to the simple Rayleigh-B{\'e}nard problem, the priciple of exchange of stabilities is not valid, and the marginal state will be characterised by non-zero frequency of oscillation.
Combining equations \eqref{eqn:reduced:vorticity} and \eqref{eqn:reduced:density} into an equation for $\varphi$, and substituting the ansatz $\varphi = \mathcal{A}\sin (m\pi x) \exp (i\gamma t + i k y)$, where $\gamma \in \mathbb{R}$, we obtain the dispersion relation  
\begin{equation}
\brac{ i \gamma \Delta_k +  \Pran \Delta_k^2 + \frac{\Lperp^2 \Omega}{\Lpar}  } \brac{i \gamma + ik\frac{\Ra^* \Pran}{\Omega}  +\Delta_k 
+ \frac{\Omega}{\Lpar}
} = 
 {{\Ra}^* \, \Pran}  \brac{ ik \frac{\Lperp^2}{\Lpar} +k^2\Delta\Theta } ,
\label{eqn:reduced:dispresion}
\end{equation}
where $\Delta_k = \brac{(m\pi)^2 +k^2}$, and $\Delta\Theta = \brac{\ln (1+\Delta n) - {2h}/{R_c}}$. 
The imaginary part of \eqref{eqn:reduced:dispresion} gives the frequency at onset:
\begin{equation}
\gamma = -k \Delta_k^2 \Pran \brac{\frac{\Ra^* \Pran}{\Omega}} 
\sqbrac{(1+\Pran)\Delta_k^2 + \frac{\Omega}{\Lpar} \brac{\Delta_k +\Lperp^2}}^{-1},
\label{eqn:reduced:frequency}
\end{equation}
and the real part gives the stability threshold: 
\begin{equation}
\Ra^*\Delta\Theta 
= \frac{1}{k^2 }
\brac{\Delta_k^2 +\frac{\Lperp^2 \Omega}{\Lpar \Pran} } 
\brac{ \Delta_k + \frac{\Omega}{\Lpar}}
+\Delta_k^3 \brac{\frac{\Ra^* \Pran}{\Omega}}^2 \frac{\Delta_k^2 + \frac{\Omega}{\Lpar} \brac{\Delta_k + \Lperp^2}}{\sqbrac{(1+\Pran)\Delta_k^2 + \frac{\Omega}{\Lpar} \brac{\Delta_k +\Lperp^2}}^2}.
\label{eqn:reduced:marginal}
\end{equation}
On inspection of expression~\eqref{eqn:reduced:marginal}, several features may be observed. First, we note that contained within expression \eqref{eqn:reduced:marginal}, though slightly obscured, is the stability threshold of convection, $\Delta_k^3/k^2$ (i.e.\ expression \eqref{eqn:RB:marginal}); it can be revealed by multiplying out the brackets in the first term on the right hand side of \eqref{eqn:reduced:marginal}. It follows, since all of the dimensionless parameters are positive, that the reduced plasma problem is more stable than the convection problem. Furthermore, unlike for the convection problem, here both the onset of instability as well as the critical wavenumber are dependent on the Prandtl number $\Pran$ (as well as on all the other parameters). Next, we observe the stabilizing effect of the compressible \EcrossB drift, defining a lower bound for the marginal stability threshold, $\ln(1+\Delta n) > 2h/R_c$, consistent with previous literature \citep{garcia2001two}. Finally, we remark on the implications of the presence of the $\Ra^{*2}$ term on the right hand side of \eqref{eqn:reduced:marginal}. Recall that in the analogous convection problem (cf. \eqref{eqn:RB:vorticity}, \eqref{eqn:RB:density}), the threshold for instability is given by expression~\eqref{eqn:RB:marginal} (where $\Ra = \Ra^* \ln(1+\Delta n)$); hence, increasing $\Ra^*$ always results in decreasing the marginal stability threshold, and thus an increasingly more unstable system. Here, on the other hand, the situation becomes more subtle: for large enough $\Ra^*$, increasing $\Ra^*$ will result in increasing the density difference at the onset, and thus a more stable system; this has also been observed by \cite{vilela2017analytical}. This stabilising effect at large $\Ra^*$ is ultimately due to the inclusion of the curvature term due to the diamagnetic drift in the density continuity equation~\eqref{eqn:density2d}. Indeed, it has been known  that interchange-driven models of SOL plasma that do no include this curvature term in the density equation become more unstable with increasing curvature drive (here represented by $\Ra^*$) \citep[e.g.][]{ghendrih2003theoretical, aydemir2005convective}.

\section{Characterisics of the instability} \label{sec:results}

\subsection{Comparison between the full and the reduced system} 

\begin{figure}
\centering
	\begin{subfigure}[t]{0.48\linewidth}
		\includegraphics[width=\linewidth]{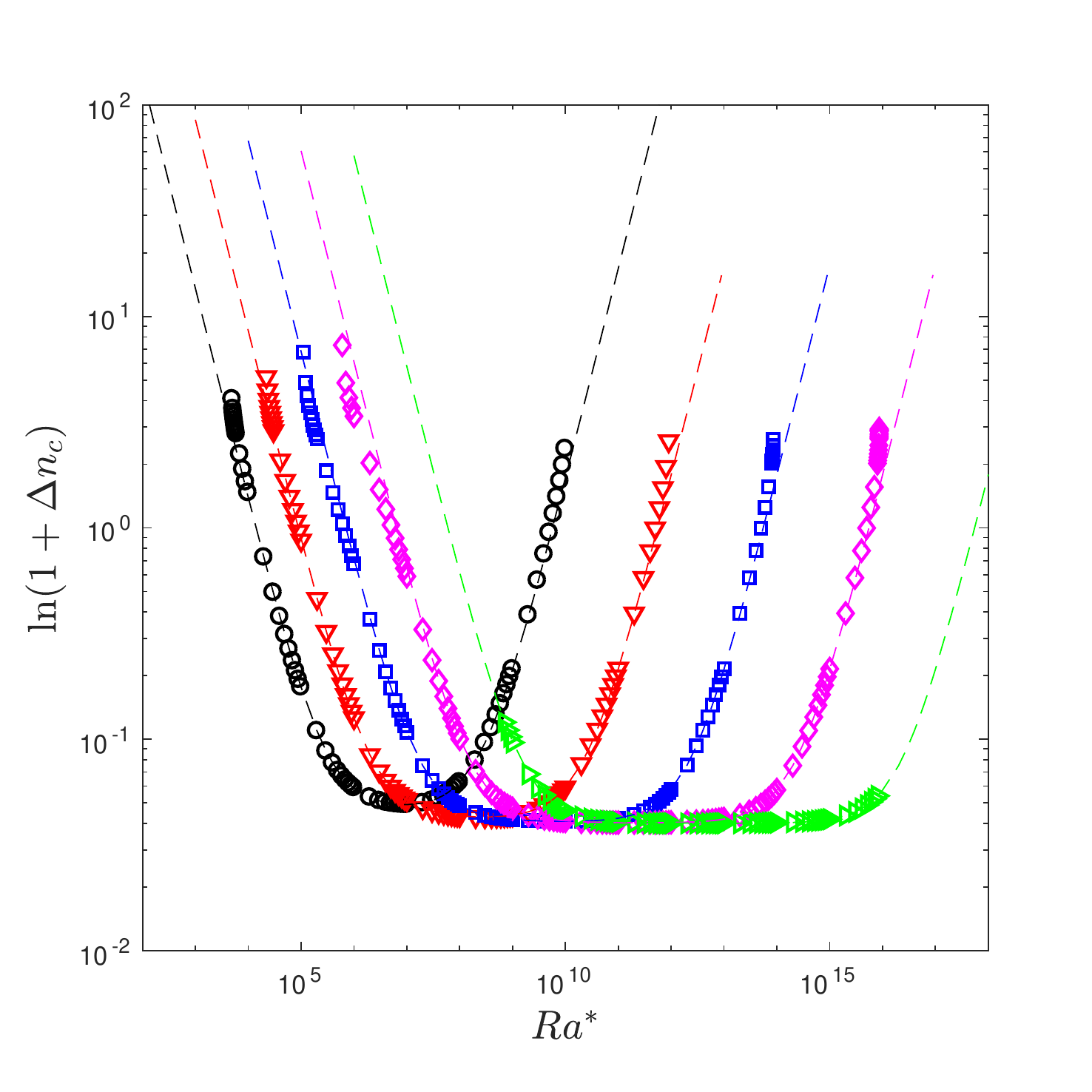}
		\caption{}
		\label{fig:mainResult_A}
	\end{subfigure}
	\begin{subfigure}[t]{0.48\linewidth}
		\includegraphics[width=\linewidth]{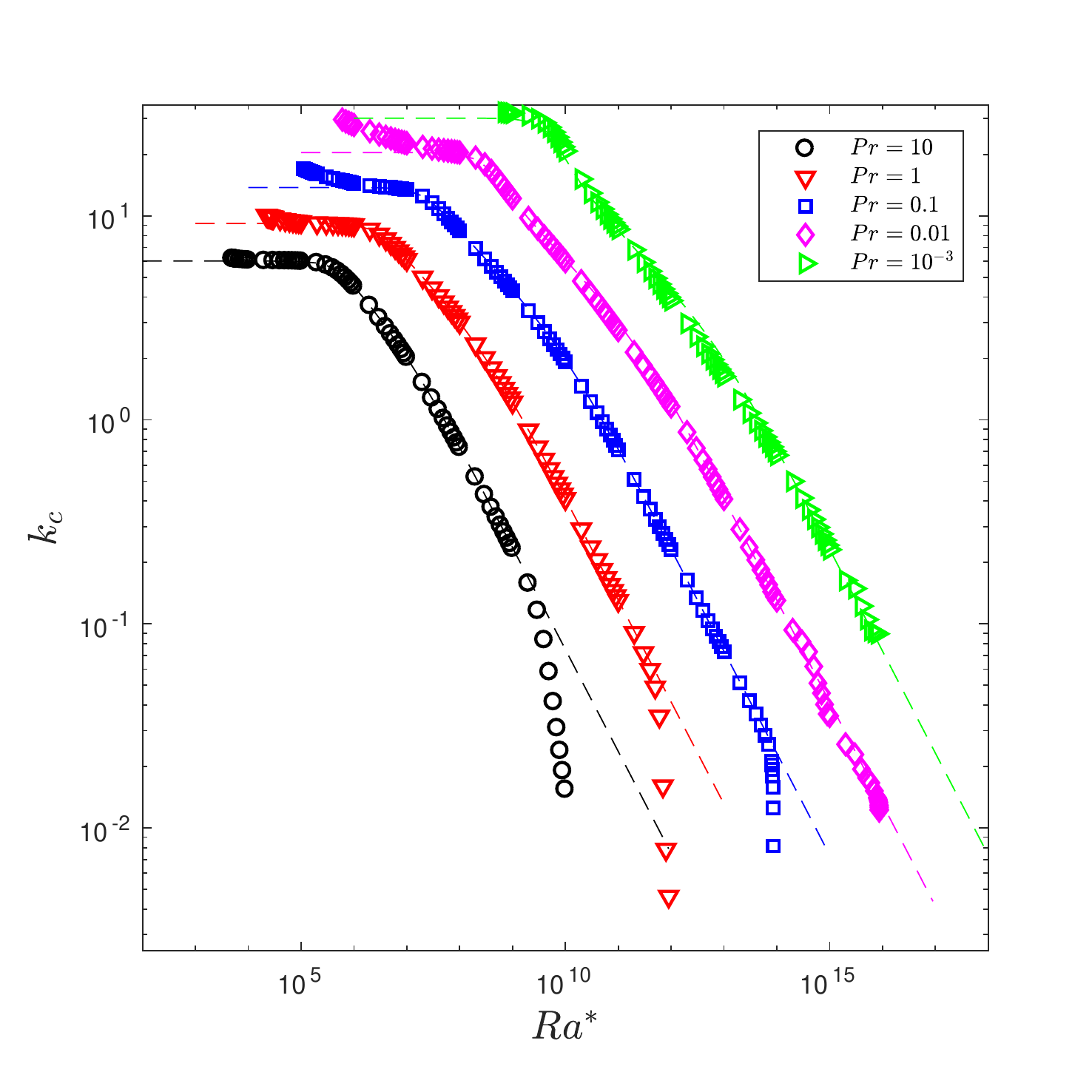}
		\caption{}
		\label{fig:mainResult_B}
	\end{subfigure}
\caption{Variation of (a)\ the critical density difference $\Delta n_c$, and (b) the corresponding critical wavenumber $k_c$, versus $\Ra^*$. Markers: full system, dashed lines: reduced system.}
\label{fig:mainResult}
\end{figure}

Figures~\ref{fig:mainResult_A} and \ref{fig:mainResult_B} show, respectively, the critical density difference and the corresponding critical wavenumber at the onset of instability. For comparison, dashed lines indicate the stability threshold and the critical $k$ of the reduced system obtained by minimising the expression for marginal stability \eqref{eqn:reduced:marginal} with respect to the wavenumber. Before comparing the two systems, let us briefly comment on the effect of varying $\Ra^*$ and $\Pran$ on the stability threshold in the full system. The critical density curves (Fig.~\ref{fig:mainResult_A}), have a roughly parabolic shape for all values of $\Pran$\,: as $\Ra^*$ is increased, $\Delta n_c$ is reduced until it reaches a minimum, after which further increase of $\Ra^*$ leads to an increase in $\Delta n_c$. Unlike in the case of Rayleigh-B{\'e}nard convection, here the onset of instability is $\Pran$ dependent. Reducing $Pr$ shifts the critical density curves to increased $\Ra^*$; i.e.\ for smaller $\Pran$, the location of the minimum of $\Delta n_c$ occurs at higher $\Ra^*$. Furthermore, the span of the trough between the two tails of each curve widens as $\Pran$ is decreased. The critical wavenumber (Fig.~\ref{fig:mainResult_B}) decreases with $\Ra^*$, but the rate at which it decreases varies with $\Ra^*$; broadly speaking the gradient of this decrease becomes steeper with increasing $\Ra^*$. Furthermore, reducing $\Pran$ for a given $\Ra^*$ increases the critical wavenumber.

\begin{figure}
\centering
	\begin{subfigure}[t]{0.48\linewidth}
		\includegraphics[width=\linewidth]{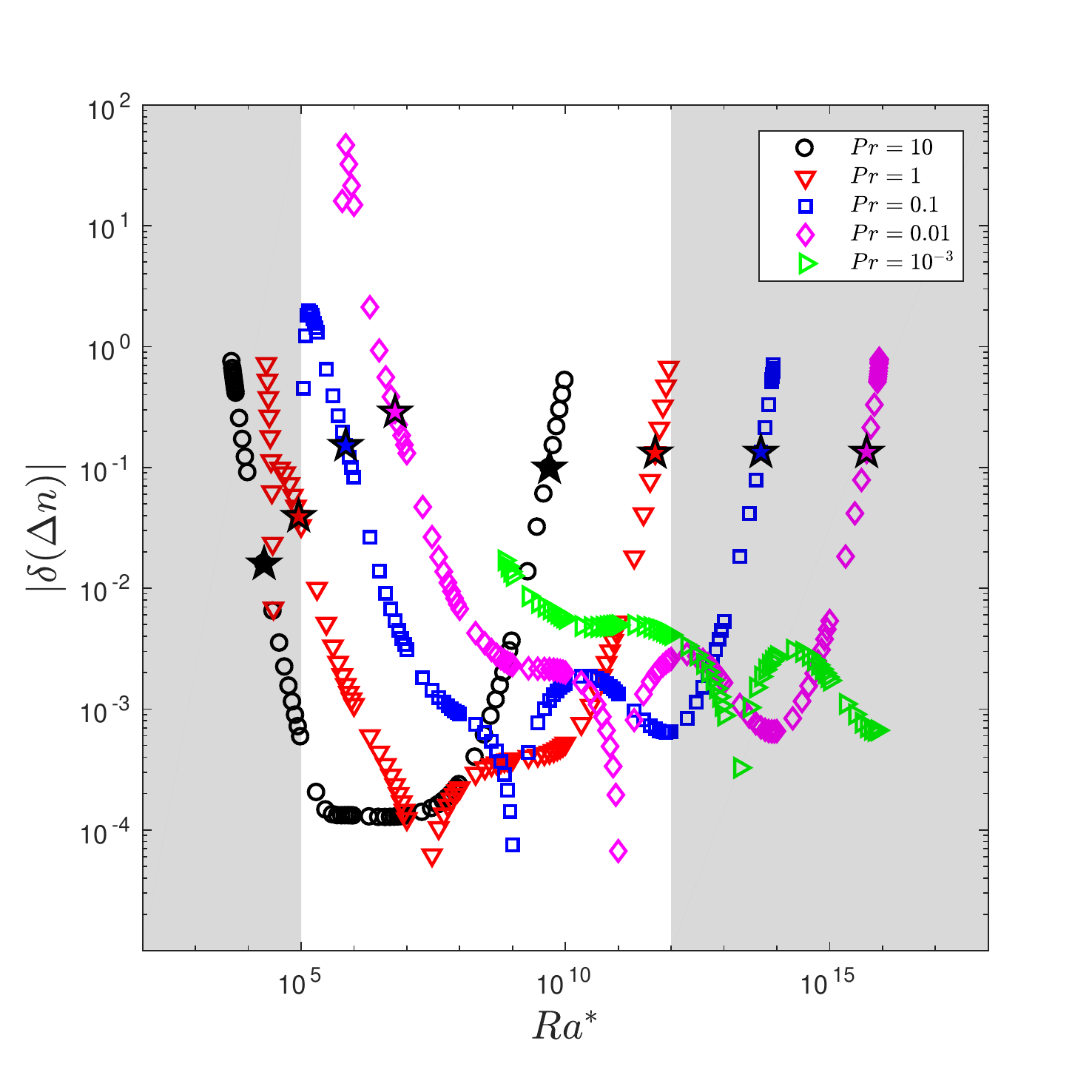}
		\caption{}
		\label{fig:relativeError_A}
	\end{subfigure}
	\begin{subfigure}[t]{0.48\linewidth}
		\includegraphics[width=\linewidth]{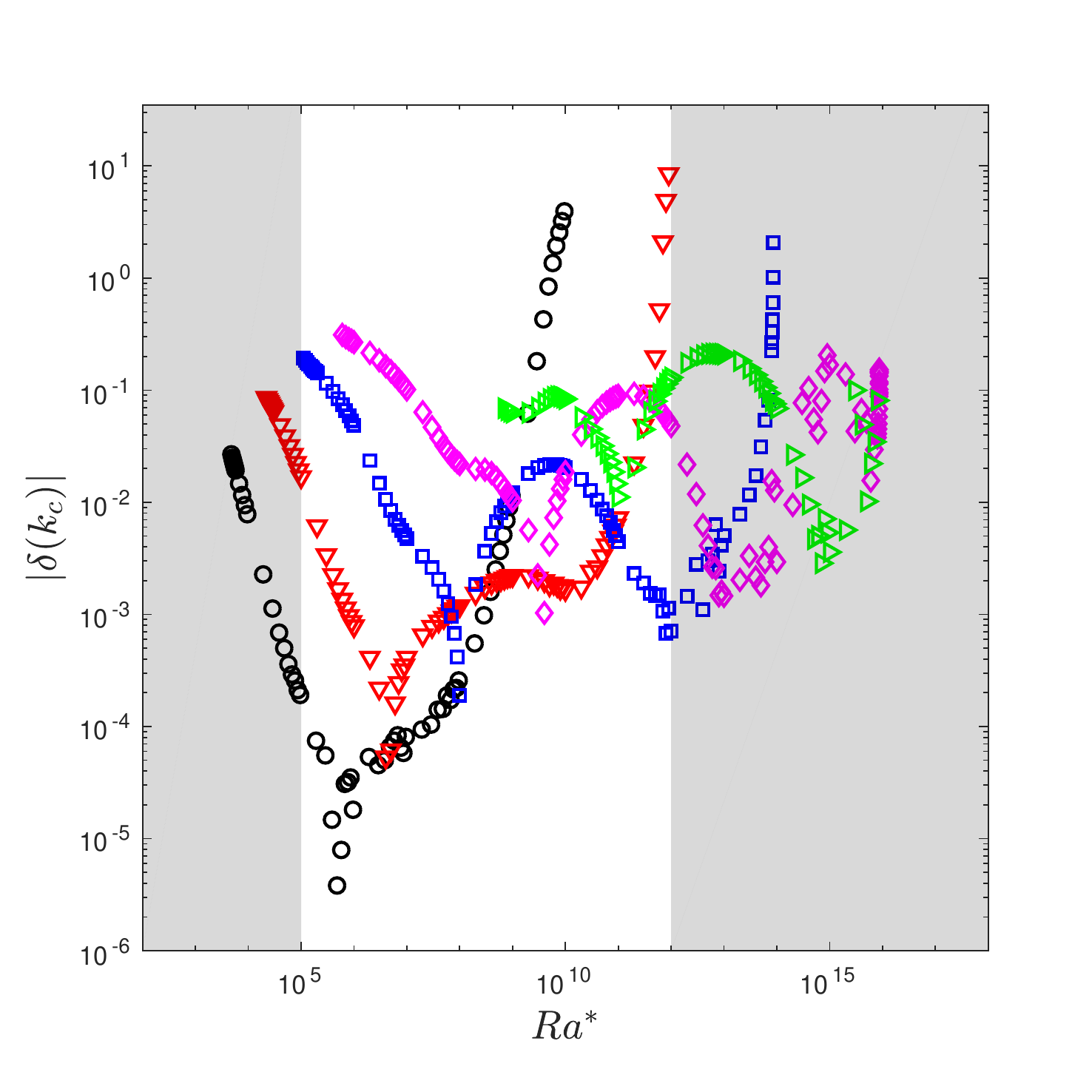}
		\caption{}
		\label{fig:relativeError_B}
	\end{subfigure}
\caption{Relative difference between the full and reduced systems for (a) the critical density difference $\Delta n_c$, and (b) the corresponding critical wavenumber $k_c$. Starred markers in (a) indicate the beginning and the end of range of $\Ra^*$ for which $\ln(1+\Delta n) < 1$. Shaded areas indicate the parameter space outside the range of interest, specified in Table \ref{table:parameters}.}
\label{fig:relativeError}
\end{figure}

\begin{figure}
\centering
	\begin{subfigure}[t]{0.33\linewidth}
		\includegraphics[width=\linewidth]{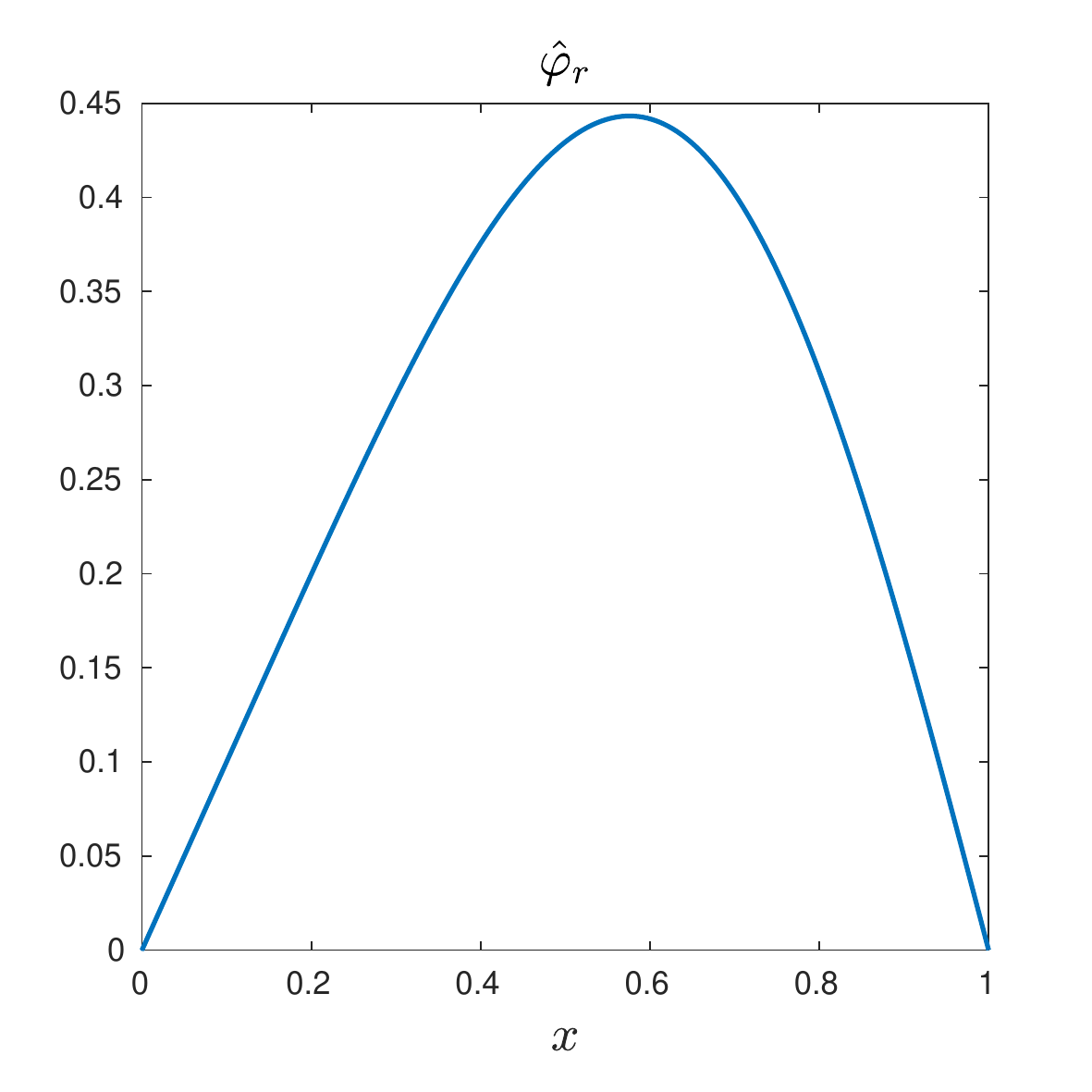}
	\end{subfigure}%
	\begin{subfigure}[t]{0.33\linewidth}
		\includegraphics[width=\linewidth]{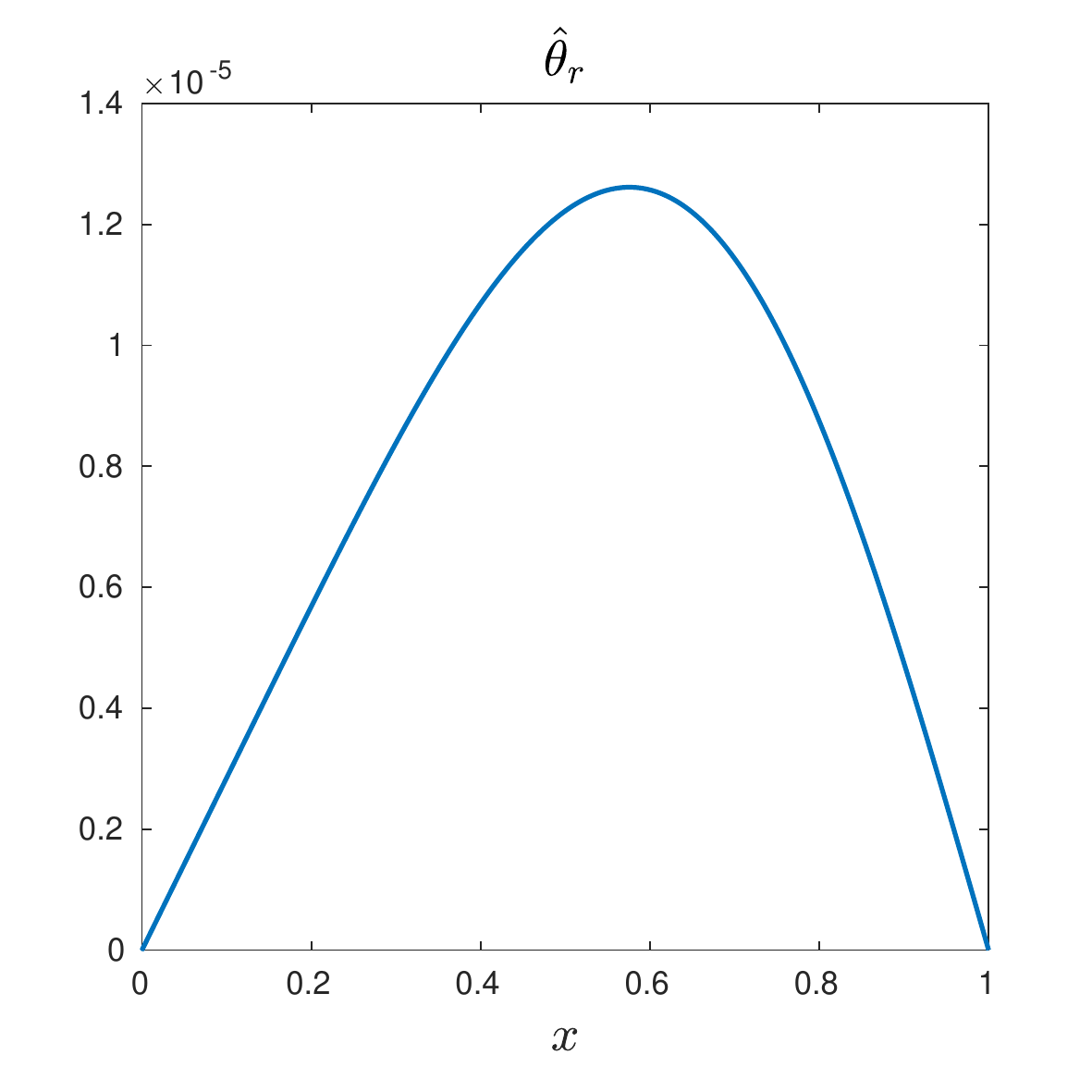}
	\end{subfigure}%
	\begin{subfigure}[t]{0.33\linewidth}
		\includegraphics[width=\linewidth]{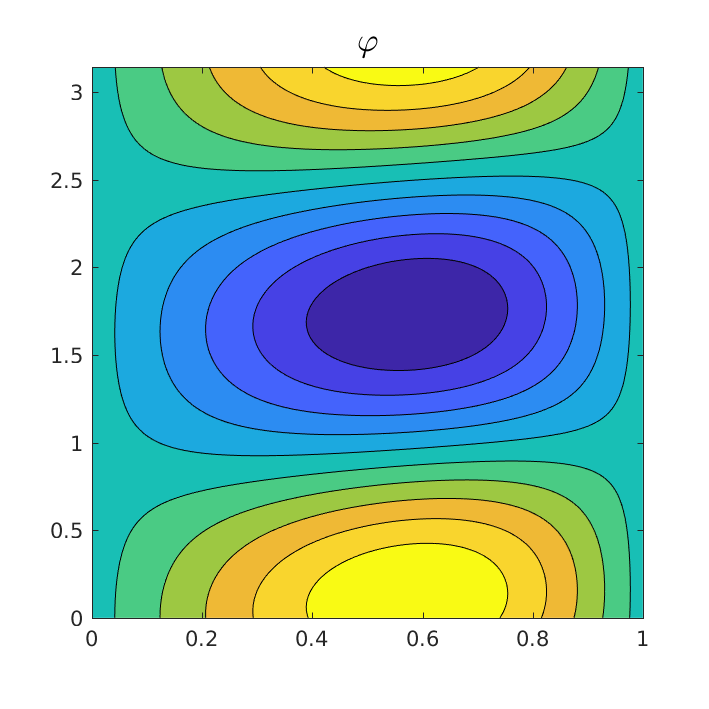}
	\end{subfigure}
\caption{Radial eigenfunction profiles $\hat{\varphi}(x)$, $\hat{\theta}(x)$, and the resulting convective cell $\varphi(x,y)$ for $\Ra^* = 10^{10}$, $\Pran = 0.1$}
\label{fig:eigfunc_modRa}
\end{figure}

Overall, we observe a remarkable agreement between the stability properties of the full system and those of the reduced system. Figure~\ref{fig:relativeError} shows the relative differences in the critical density difference and the critical wavenumber in the two systems. For each value of $\Pran$, there exists a range of $\Ra^*$ values within which the critical density gradient and critical wavenumber of the reduced system are good approximations to their counterparts in the full system. These regions of agreement are characterised by low values of $\Delta n_c$ and therefore a basic state gradient that is close to uniform. In these circumstances, the radial structure of the eigenfunctions of the full system, shown in Figure~\ref{fig:eigfunc_modRa}, closely resembles that of the sinusoidal solutions of the reduced, constant coefficient system. Outside the ranges of agreement, the inhomogeneity of the basic state gradient becomes more pronounced, with the terms that were ignored in constructing the reduced system becoming significant in influencing the stability threshold and the structure of the solutions. In particular, when $\Delta n_c$ becomes very large, the eigenfunctions develop sharp gradients near $x = 1$, characteristic of a boundary layer problem. This will be discussed in more detail below (see section \ref{sec:boundary_layer}).

Perhaps unsurprisingly, the agreement between the full and reduced systems begins to break down when $\Delta n_c$ grows to order unity, or, equivalently, in the limits of small and large $\Ra^*$, though the precise meaning of `small' and `large' $\Ra^*$ is dependent on the value of $\Pran$. We shall attempt to elucidate those meanings in the following sections. In both the full and the reduced systems, the critical density difference $\Delta n_c$ grows indefinitely in the limits of small and large $\Ra^*$. For small $\Ra^*$, the critical wavenumber in the reduced system tends to a constant value that is dependent on $\Pran$. In the full system, on the other hand, as $\Ra^*$ is reduced, $k_c$ appears to grow indefinitely. In the opposite limit, as $\Ra^*$ is increased, the critical wavenumber decreases towards zero faster in the full system than in the reduced system.


\subsection{Properties of the reduced system}
In this section we study the variation of the critical value of $\Delta n$ and of the corresponding critical $k$ with $\Ra^*$ in the reduced system. Through this investigation we will attempt to estimate the range of $\Ra^*$ for which there is good agreement between the stability threshold of the full and the reduced systems. As observed above, agreement between the two systems is good provided that the density difference is small, in particular when $\ln (1+\Delta n_c) < 1$. By considering the behaviour of the critical density threshold in the reduced system, we can approximate the values of $Ra^*$ at which the agreement breaks down, i.e.\ when $\ln (1+ \Delta n_c)$ exceeds unity.

Recall that the critical density difference $\Delta n_c$ in the reduced system is obtained by minimising \eqref{eqn:reduced:marginal} with respect to $k$. On differentiating \eqref{eqn:reduced:marginal} with respect to $k^2$ and setting the result to zero, we find that the critical wavenumber satisfies the following equation: 
\begin{align}
\left\lbrace(2k^2 - \pi^2)\Delta_k^2 + \frac{\Omega}{\Lpar} (k^2 - \pi^2)\Delta_k - \pi^2 \frac{\Lperp^2 \Omega}{\Lpar \Pran} - \frac{1}{\Pran}  \brac{\frac{\Lperp \Omega}{\Lpar}}^2 \right\rbrace
+ k^4 \brac{\frac{\Ra^* Pr}{\Omega}}^2 \xi\brac{k, \Pran,\Omega, \Lperp, \Lpar} = 0, \label{eqn:crit_k}
\end{align}
where
\begin{align}
\xi\brac{k, \Pran,\Omega, \Lperp, \Lpar} = \frac{\splitdfrac{\Delta_k^2 \sqbrac{(1+\Pran)\Delta_k^2 + \frac{\Omega}{\Lpar} (\Delta_k +\Lperp^2)} \sqbrac{5 \Delta_k^2 + \frac{\Omega}{\Lpar} (4 \Delta_k + 3 \Lperp^2)} \ldots }{\hspace{40mm} - 2\Delta_k^3 \sqbrac{2(1+\Pran)\Delta_k + \frac{\Omega}{\Lpar}}\sqbrac{\Delta_k^2 + \frac{\Omega}{\Lpar}(\Delta_k + \Lperp^2)}}}{\sqbrac{(1+\Pran)\Delta_k^2 + \frac{\Omega}{\Lpar} (\Delta_k +\Lperp^2)}^3}.
\label{eqn:crit_k2}
\end{align}
Attempting to extract an analytical solution for $k_c$ is a hopeless task, but we can gain useful insight by considering the limits of small and large $\Ra^*$.

\subsubsection{Behaviour at small $\Ra^*$}

\begin{figure}
\centering
	\begin{subfigure}[t]{0.48\linewidth}
		\includegraphics[width=\linewidth]{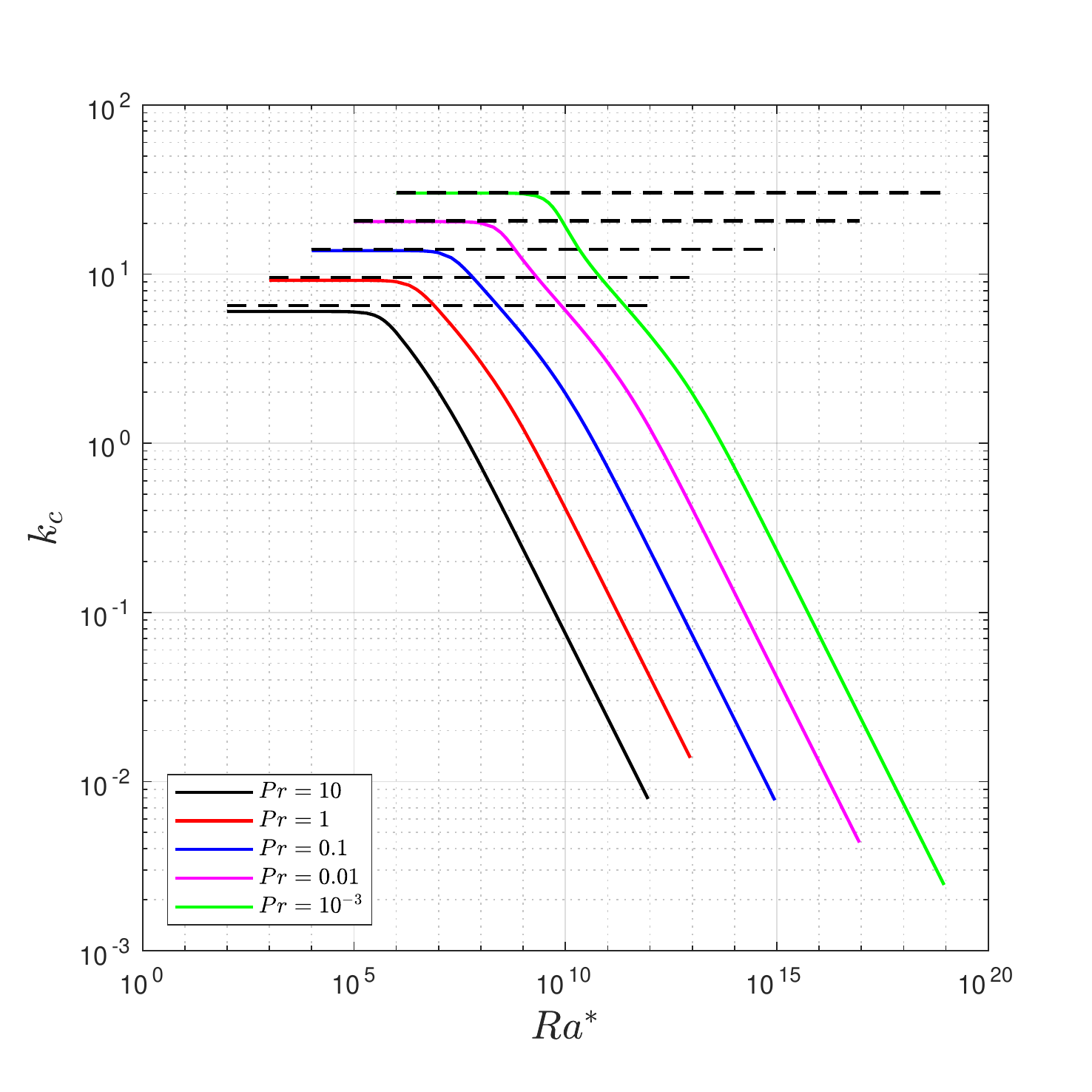}
		\caption{}
		\label{fig:RSkc1}
	\end{subfigure}
	\begin{subfigure}[t]{0.48\linewidth}
		\includegraphics[width=\linewidth]{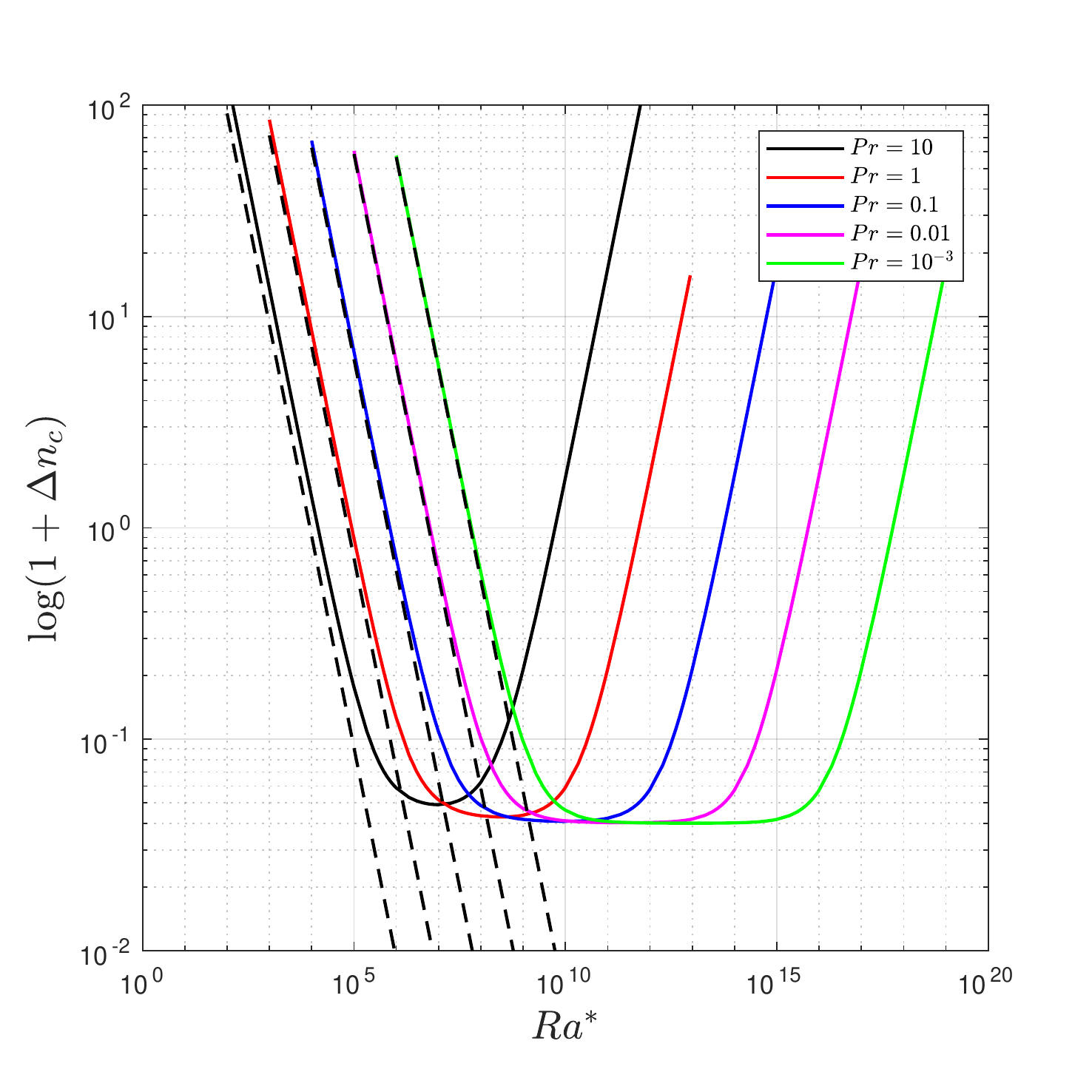}
		\caption{}
		\label{fig:RSdn1}
	\end{subfigure}
\caption{Scaling behaviour at small $\Ra^*$. (a) Variation of $k_c$ as given by solution of \eqref{eqn:crit_k} (solid lines) and the approximation \eqref{eqn:kc_lowRa:scaling} (dashed lines). (b) Variation of $\Delta n_c$ given by solution of \eqref{eqn:reduced:marginal} (solid lines) and the approximation \eqref{eqn:dnc_lowRa:scaling} (dashed lines).}
\end{figure}

When the factor $(\Ra^* \Pran/\Omega)^2$ in equation \eqref{eqn:crit_k} is sufficiently small, the critical value  $k_c$ becomes independent of $\Ra^*$, and can be approximated by a solution of 
\begin{equation}
(2k^2 - \pi^2)\Delta_k^2 + \frac{\Omega}{\Lpar} (k^2 - \pi^2)\Delta_k - \pi^2 \frac{\Lperp^2 \Omega}{\Lpar \Pran} - \frac{1}{\Pran}  \brac{\frac{\Lperp \Omega}{\Lpar}}^2 = 0.
\label{eqn:kc_lowRa}
\end{equation}
Provided that $\Omega/\Pran$ is large (which is to be expected), dominant balance dictates that
\begin{equation}
k_c^6 \sim \frac{1}{2 \Pran} \frac{\Lperp^2 \Omega}{\Lpar}\brac{\pi^2 + \frac{\Omega}{\Lpar}}.
\label{eqn:kc_lowRa:scaling}
\end{equation}
Figure \ref{fig:RSkc1} shows that the dominant balance estimate for $k_c$ and the true solution of \eqref{eqn:crit_k} are in good agreement, and that the agreement improves for smaller values of $\Pran$. We note also that for each value of $\Pran$ the true solution curves begin to deviate from the approximations only when the factor $(\Ra^* \Pran/\Omega)^2$ grows to $\mathcal{O} (10^2)$. Using \eqref{eqn:kc_lowRa:scaling} in \eqref{eqn:reduced:marginal} gives the following scaling for the critical density difference $\Delta n_c$ in the small $\Ra^*$ limit,
\begin{equation}
\Ra^* \log(1+\Delta n_c) \sim \frac{ 1 }{\Pran} \brac{\frac{\Lperp^2 \Omega}{\Lpar}} +  \frac{ \sqrt[3]{2} }{\Pran^{2/3}}\brac{\frac{\Lperp^2 \Omega}{\Lpar}\brac{\pi^2 + \frac{\Omega}{\Lpar}}}^{2/3}.
\label{eqn:dnc_lowRa:scaling}
\end{equation}
This estimate is compared to the true variation of $\Delta n_c$ in Figure \ref{fig:RSdn1}; again we observe good agreement between the two results, especially when $\Pran$ is small.
For completeness, we need to provide some (appropriate) interpretation of what ``small'' $\Ra^*$ means in this context. In this matter we adopt a pragmatic approach. Bearing in mind that we are interested in conditions for which $\ln(1 +\Delta n_c) \approx 1$, and observing that the curves and the scalings in Figure \ref{fig:RSdn1} indeed align when $\ln (1+\Delta n)$ is above unity, from \eqref{eqn:dnc_lowRa:scaling} we estimate that $\Ra^*$ can be considered ``small'' if
\begin{equation}
\Ra^* \lesssim \frac{ 1 }{\Pran} \brac{\frac{\Lperp^2 \Omega}{\Lpar}} +  \frac{ \sqrt[3]{2} }{\Pran^{2/3}}\brac{\frac{\Lperp^2 \Omega}{\Lpar}\brac{\pi^2 + \frac{\Omega}{\Lpar}}}^{2/3}.
\end{equation}

\subsubsection{Behaviour at large $\Ra^*$}

\begin{figure}[h!]
\centering
	\begin{subfigure}[t]{0.5\linewidth}
		\includegraphics[width=\linewidth]{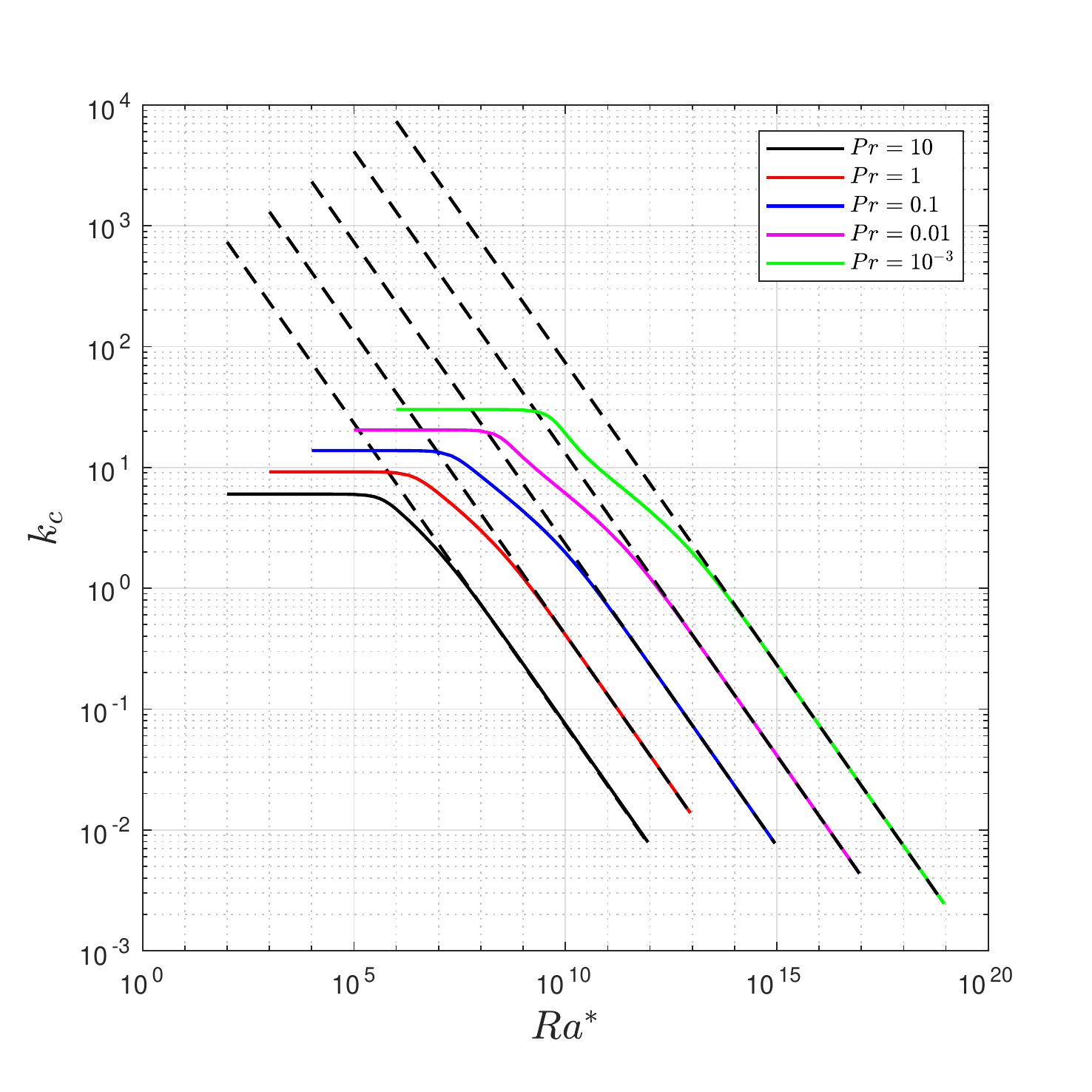}
		\caption{}
		\label{fig:RSkc2}
	\end{subfigure}%
	\begin{subfigure}[t]{0.5\linewidth}
		\includegraphics[width=\linewidth]{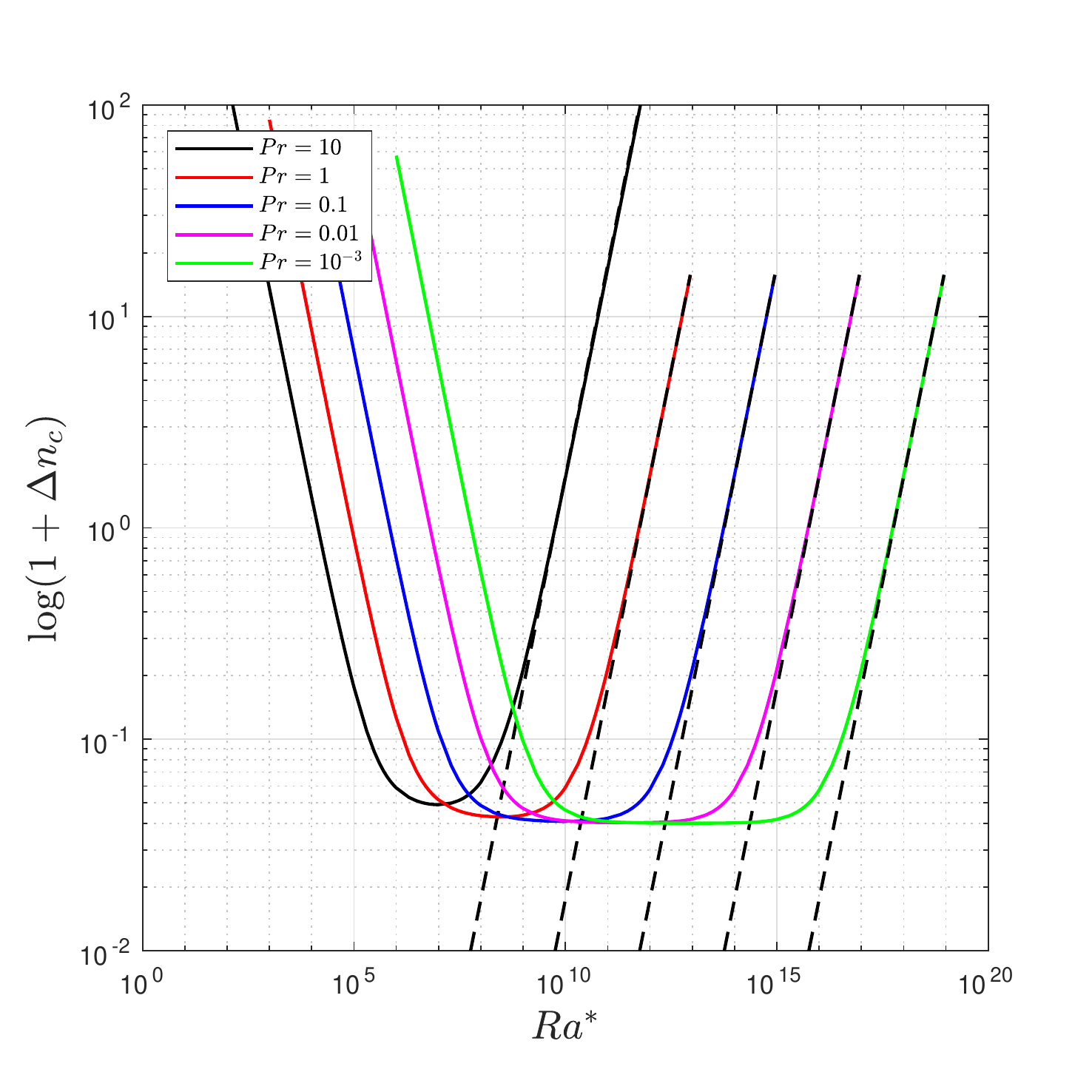}
		\caption{}
		\label{fig:RSdn2}
	\end{subfigure}
\caption{Scaling behaviour at large $\Ra^*$. (a) Variation of $k_c$ as given by solution of \eqref{eqn:crit_k} (solid lines) and the approximation \eqref{eqn:kc_largeRa:scaling} (dashed lines). (b) Variation of $\Delta n_c$ given by solution of \eqref{eqn:reduced:marginal} (solid lines) and the approximation \eqref{eqn:dnc_largeRa:scaling} (dashed lines).}
\end{figure}

When $\Ra^*$ becomes large, the last term in \eqref{eqn:crit_k} would appear to dominate, suggesting that $k_c$ is a root of $\xi\brac{k, \Pran,\Omega, \Lperp, \Lpar} =0$, with $\xi$ defined in \eqref{eqn:crit_k2}. However, we found that for the parameter values under consideration, this equation has no roots. Consequently, the large $\Ra^{*2}$ term in \eqref{eqn:crit_k} has to be balanced by other terms in that equation. We therefore expect the dominant balance to be given by 
\begin{equation}
\frac{\pi^2}{\Pran}\frac{\Lperp^2 \Omega}{\Lpar} + \frac{1}{\Pran}  \brac{\frac{\Lperp \Omega}{\Lpar}}^2 \sim 3 \pi^4 k^4 \brac{\frac{\Ra^* Pr}{\Omega}}^2 \brac{\frac{\Lperp^2 \Omega}{\Lpar}}^{-1},
\end{equation}
which leads to the scaling
\begin{equation}
k_c \sim \brac{\frac{\Lperp^4 \Omega^4 }{3 \pi^4 \Lpar^2}\brac{\pi^2 +\frac{\Omega}{\Lpar}}}^{1/4} \Ra^{*-1/2} \Pran^{-3/4}.
\label{eqn:kc_largeRa:scaling}
\end{equation}
Using this estimate for $k_c$ in \eqref{eqn:reduced:marginal} gives the scaling for the critical density $\Delta n_c$ at large $\Ra^*$,
\begin{equation}
\log(1+\Delta n_c) \sim  \frac{\pi^6 \Lpar}{\Lperp^2 \Omega^3} \Ra^* \Pran^2 .
\label{eqn:dnc_largeRa:scaling}
\end{equation}
Figures \ref{fig:RSkc2} and \ref{fig:RSdn2} compare estimates~\eqref{eqn:kc_largeRa:scaling} and \eqref{eqn:dnc_largeRa:scaling} with the true values obtained from numerical solution of the full system~\eqref{eqn:eigenvalue_system}. In both cases the agreement is remarkable given the simplicity of the scalings and the complexity of the true solution. 
As for the case of small $Ra^*$, we again observe that the true $\Delta n_c$ curves match their respective scalings when $\log(1+ \Delta n_c)$ exceeds unity. We may therefore estimate from \eqref{eqn:dnc_largeRa:scaling} that the ``large'' $\Ra^*$ regime is defined by
\begin{equation}
\Ra^* \gtrsim \frac{\Lperp^2 \Omega^3}{\pi^6 \Lpar}  \Pran^{-2} .
\end{equation}

\subsubsection{Implications for the behaviour of the full system}
Once we have identified the regions of small and large $\Ra^*$, can we safely say that outside these regions the reduced system is a good approximation to the full system? The best answer we can offer is `tentatively'. We would certainly expect the agreement to break down when the terms unaccounted for by the reduced system (related to $\Delta n$) grow in magnitude to order unity (i.e.\ when $\ln (1+\Delta n) > 1$). Furthermore, the relative error between the stability boundaries in the two systems remains below acceptable levels within these ranges (recall Figure \ref{fig:relativeError_A}). However, we cannot ignore the general trend of the error curves in Figure \ref{fig:relativeError_A} to shift upwards as $\Pran$ is reduced. It is plausible that for values of $\Pran$ smaller than those investigated here, the window of agreement between the two systems could shrink. Nonetheless, for the range of $\Pran$ values of interest (see Table \ref{table:parameters}), the estimates of small and large $\Ra^*$ can be used as approximate lower and upper bounds of $\Ra^*$ between which the reduced system is a good predictor of the behaviour in the full system. Finally, we can comment briefly on the behaviour of the full system in the limit of large $\Ra^*$. Although we are not able to extract any precise scaling for the behaviour of the critical wavenumber in the full system, we can conclude, by comparison with the reduced system, that for large $\Ra^*$, $k_c$ decays faster than $\Ra^{*-1/2}$. Furthermore, the rate of this decay increases as $\Pran$ is decreased.

\subsection{Beyond the reduced system - a boundary layer problem} \label{sec:boundary_layer}
\begin{figure}
\centering
	\begin{subfigure}[t]{0.33\linewidth}
		\includegraphics[width=\linewidth]{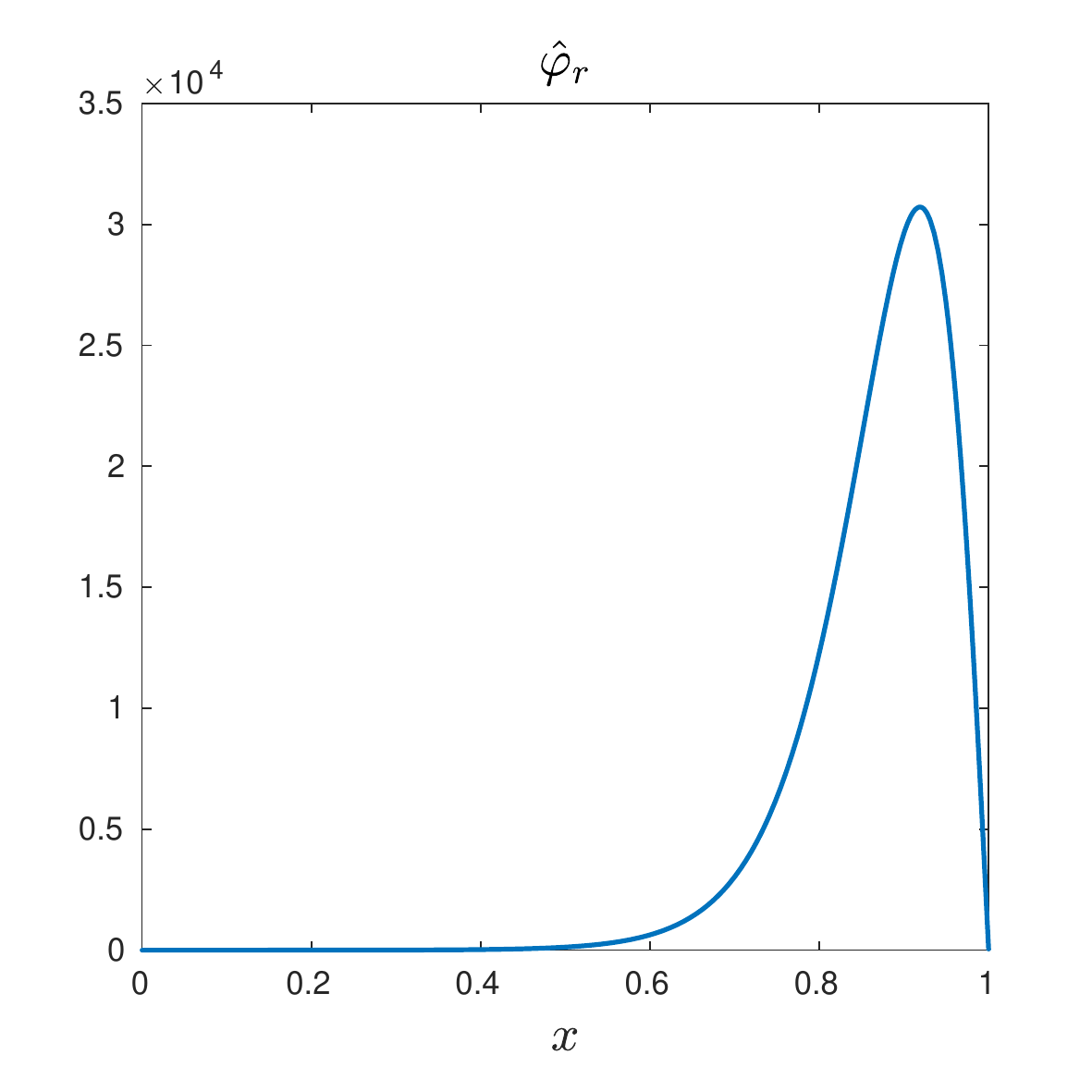}
	\end{subfigure}%
	\begin{subfigure}[t]{0.33\linewidth}
		\includegraphics[width=\linewidth]{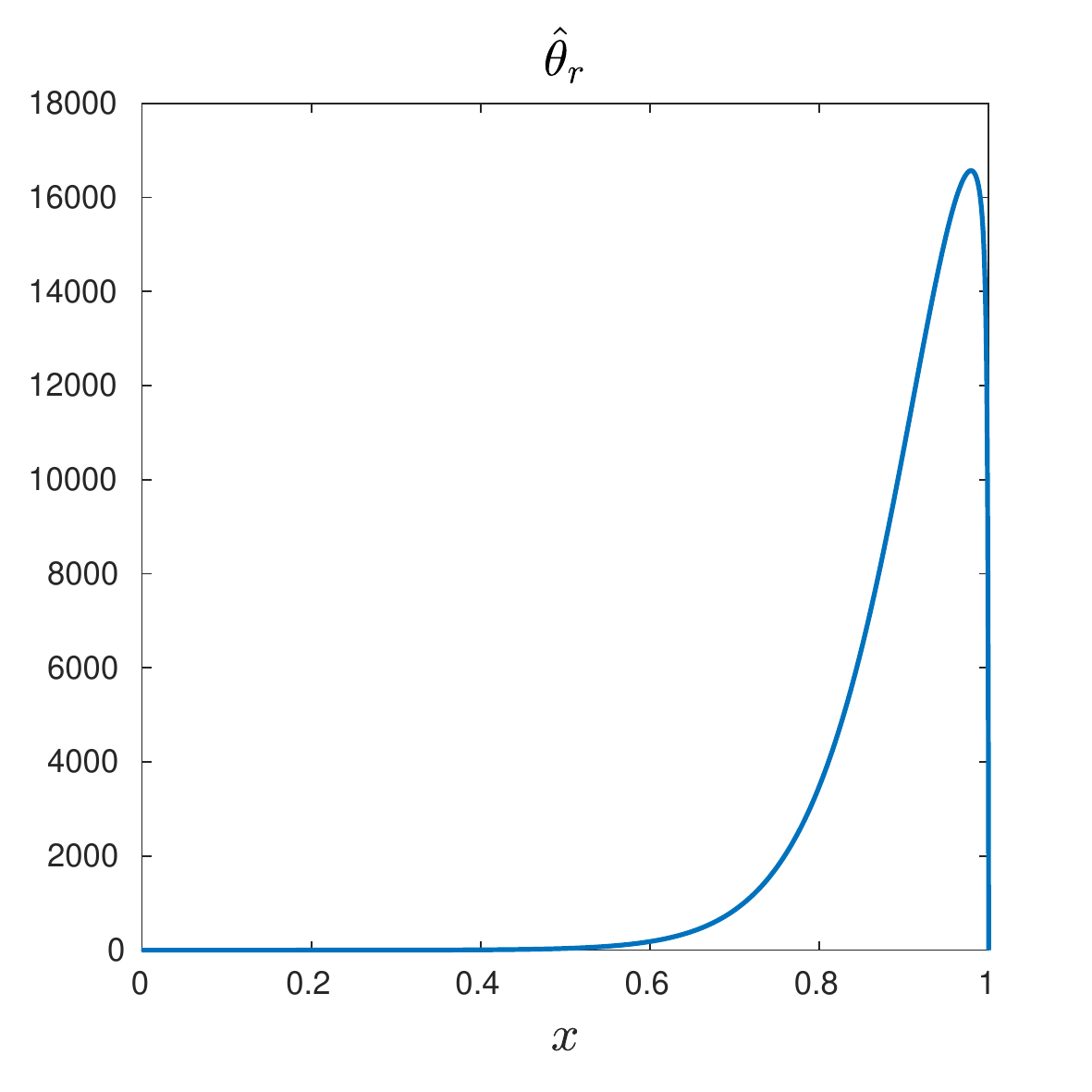}
	\end{subfigure}%
	\begin{subfigure}[t]{0.33\linewidth}
		\includegraphics[width=\linewidth]{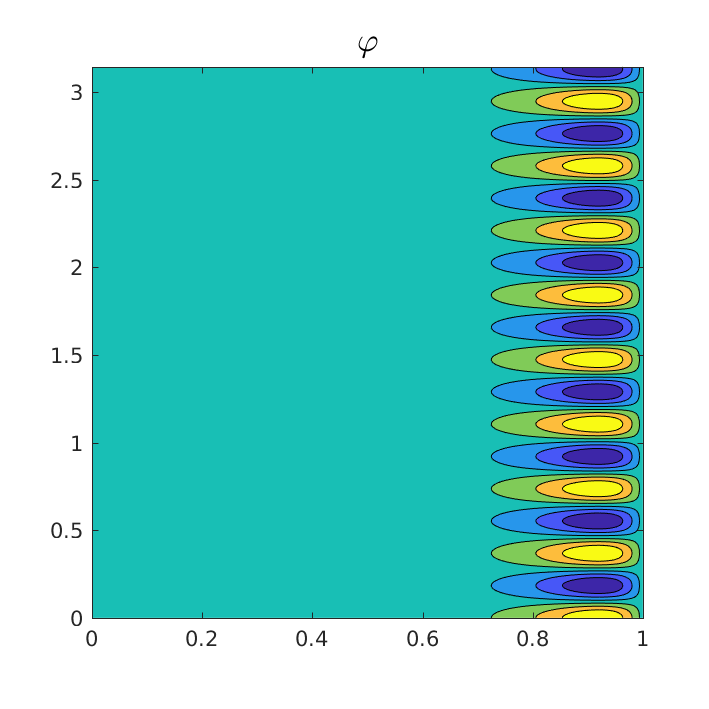}
	\end{subfigure}
\caption{Radial eigenfunction profiles $\hat{\varphi}(x)$, $\hat{\theta}(x)$, and the resulting convective cell $\varphi(x,y)$ for $\Ra^* = 2\times 10^{5}$, $\Pran = 0.1$}
\label{fig:eigfunc_lowRa}
\end{figure}
Recall from Figure \ref{fig:mainResult_A} that $\Delta n_c$ increases indefinitely in the limits of very small or very large $\Ra^*$. As $\Delta n_c$ grows, the inhomogeneity of the basic state gradient becomes more marked. In particular, as $\Delta n_c \rightarrow \infty$, $\Theta' \rightarrow -(1-x)^{-1}$: thus the basic state gradient develops a singularity at $x = 1$. Consequently, the $\varphi$ and $\theta$ eigenfunctions develop sharp gradients near $x = 1$, as shown in Figure~\ref{fig:eigfunc_lowRa}. This behaviour is characteristic of a boundary layer problem. Although solution of the boundary layer problem lies beyond the scope of this paper, here we demonstrate, through fairly simple means, that it is possible to extract the inner boundary layer solution, and to verify that it is consistent with numerically obtained solutions of the full system.

\begin{figure}
\centering
\includegraphics[width=0.5\linewidth]{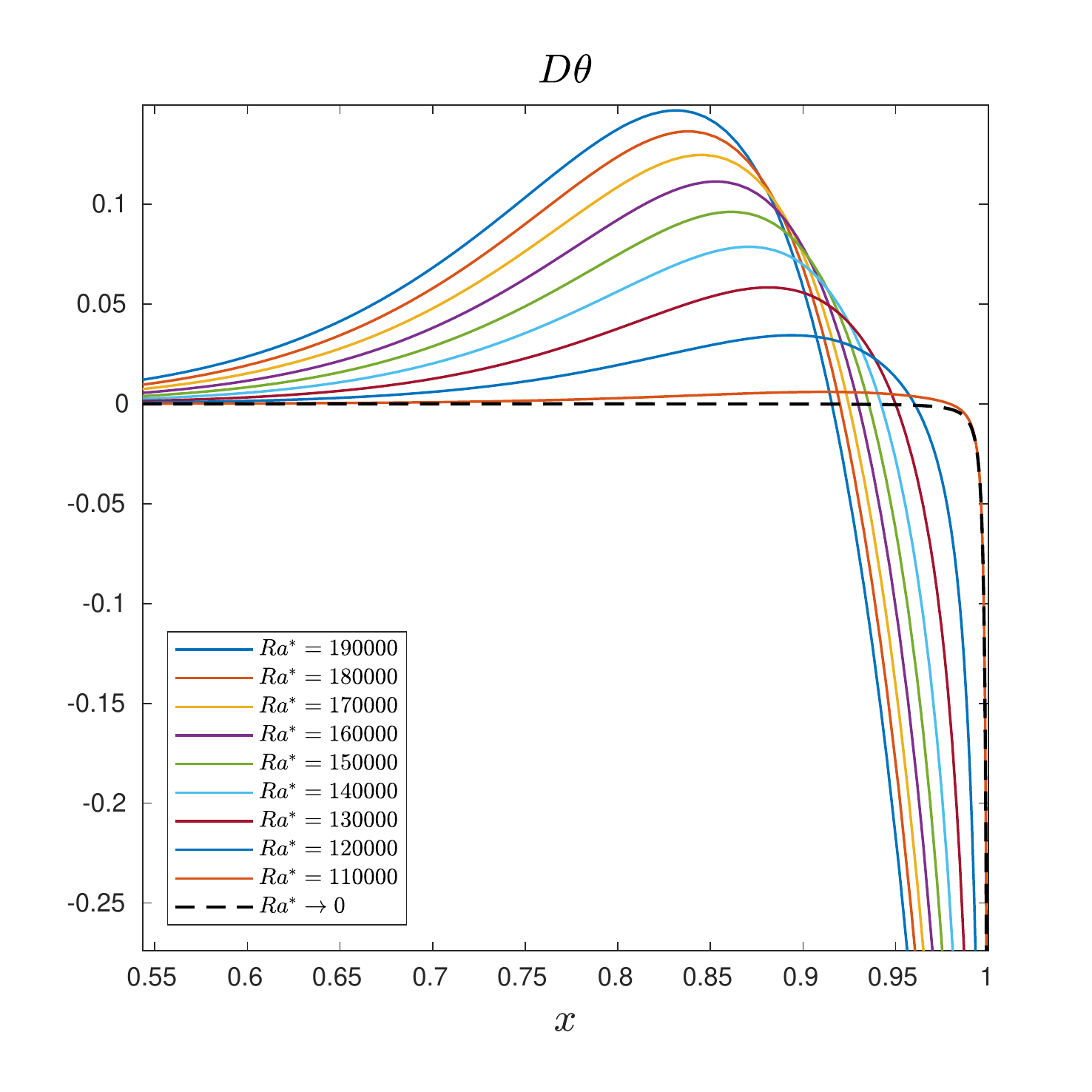}
\caption{Boundary layer behaviour at small $\Ra^*$, $\Pran = 0.1$. Solid lines are numerical solutions of $\mathcal{D}\theta$ for a range of decreasing $\Ra^*$. The dashed line is the inner boundary layer solution \eqref{eqn:bl:dtheta} (evaluated with $\Delta n_c$ matching that of the numerical solution with lowest $\Ra^*$). All profiles have been normalised by $\mathcal{D}\theta(x=1)$. }
\label{fig:bl_dtheta}
\end{figure}

We consider the log density equation \eqref{eqn:density:linear} expressed in normal mode form:
\begin{gather}
\sigma \theta= 
\brac{\dev{\Theta}{x} + \frac{2 h}{ R_c} } i k \varphi
- i \frac{Ra^* \, Pr}{ \, \Omega} k \theta +  (\D^2 \theta - k^2 \theta) + 2\dev{\Theta}{x} \D\theta 
- \frac{\Omega}{L_\parallel} \theta
+ \frac{L_\perp^2}{L_\parallel} \, \varphi .
\label{eqn:bl:thetaEquation}
\end{gather}
Suppose that there is a boundary layer near $x = 1$, where gradients in $x$ are large. Let $\varepsilon \ll 1$ be the ordering parameter, with the only ordering assumption being that derivatives in $x$ are large, with $\mathrm{d}/\mathrm{d} x \sim \mathcal{O}({1/\varepsilon})$. It then follows from \eqref{eqn:bl:thetaEquation} that the dominant balance inside the boundary layer, at $\mathcal{O}\brac{1/\varepsilon^2}$, is governed by the ordinary differential equation: 
\begin{equation}
0 = \D^2 \theta_{in} + 2\dev{\Theta}{x} \D\theta_{in}.
\end{equation}
This can be integrated to obtain
\begin{equation}
\D\theta_{in} = A \exp(-2\Theta(x))= A \brac{1+\Delta n (1-x)}^{-2}.
\label{eqn:bl:dtheta}
\end{equation}
In Figure~\ref{fig:bl_dtheta}, we compare numerically obtained solutions for $\mathcal{D}\theta$ to the proposed inner boundary layer solution \eqref{eqn:bl:dtheta} for a range of decreasing $\Ra^*$ . It can be seen that the numerical solutions tend to the profile given by \eqref{eqn:bl:dtheta}. This provides evidence that for very large $\Delta n$, the linear SOL equations have the nature of a boundary layer problem. 
\section{Discussion} \label{sec:discussion}
In this article, we have pursued two closely related objectives. The first is an in-depth linear stability analysis of a two-dimensional fluid model often used to study SOL dynamics. In this regard, we focus on characterising the conditions at the onset of instability. Specifically, we calculate the stability threshold and investigate its dependence on various plasma parameters.

At the same time we revisit, and explore further, the analogy between the SOL plasma problem and Rayleigh-B{\'e}nard convection in neutral fluids. In this respect, we demonstrate that the SOL plasma equations can indeed be reduced to those describing thermal convection with additional effects, in which analogues of the dimensionless Rayleigh and Prandtl numbers can be identified. The presence of these additional terms, however, makes the analogy not entirely straightforward: indeed the SOL stability problem differs markedly from that of Rayleigh-B{\'e}nard convection in three important respects.

First, the Rayleigh number $\Ra^*$ makes an explicit appearance in two terms in the SOL system. One is in the interchange drive term in the vorticity equation, which is a direct analogue of the buoyancy term responsible for driving the instability in the case of thermal convection; this term is therefore understood to be destabilising. In the convection problem, increasing $\Ra^*$ will result in lowering the critical density threshold and thus an increasingly more unstable system. In the SOL problem, $\Ra^*$ also appears in the density continuity equation in the term representing density flux due to diamagnetic drift; this term has a stabilising effect, and will compete with the destabilising effect of the interchange drive term. Overall, increasing $\Ra^*$ will initially have a destabilising effect up to a certain point, beyond which any further increase in $\Ra^*$ will be stabilising.  

Second, we observe that the stability threshold is Prandtl number dependent, unlike in the case of RBC. As can be seen in Figure \ref{fig:mainResult_A}, this dependence is not straightforward: at small enough $\Ra^*$ the critical density difference required for the onset of instability decreases with increasing $\Pran$, whereas at large $\Ra^*$ this trend is reversed.

Third, the basic state log density gradient in the SOL problem is non-uniform; as a result, the equations contain coefficients with explicit $x$ dependence. In contrast to the convection problem, which for idealised boundary conditions can be solved exactly, the presence of non-constant coefficients in the plasma problem makes it impossible to extract an analytical expression for the marginal stability threshold; in general, the problem has to be tackled numerically. To make analytical progress, the background gradient is sometimes approximated by a constant value (for example, \cite{vilela2017analytical} represent the gradient by the inverse of the scale length for the exponential decay of density in the SOL). Similarly here, we also consider a simplified constant-coefficient ordinary differential equation, which can be solved in exactly the same way as for the convection problem. This reduced system provides useful insight into the qualitative behaviour of the full problem, accurately predicting the responses of $\Delta n_c$ and $k_c$ to variations in $\Ra^*$ and $\Pran$. Furthermore, for each value of $\Pran$ we have identified an approximate range of $\Ra^*$ for which there is good quantitative agreement between the full and reduced systems. Outside the regions of agreement, the full system exhibits complex behaviour that cannot be explained by the simplified system. In particular, we demonstrated that owing to the spatial dependence of the background gradient, the radial structure of the solutions of the linear system can become highly localised, to the point of developing a boundary layer.

The analysis included in this paper has been guided by the long term motivation of uncovering the mechanism for the generation of plasma filaments at the edge of magnetic confinement devices. In the current work we have elucidated the analogy between the simple SOL plasma models and Rayleigh-B{\'e}nard convection, and have thereby gained insight into the fundamental stability problem. As such, this study constitutes a successful first step towards the long term goal and also paves the way for analytical considerations of more complicated models. Since it is believed that filaments are generated in the core region before being ejected to the scrape-off layer, the natural step for extending the current work is the consideration of a two-layer model in which the domain encompasses both of those regions.
\section*{Acknowledgements}
This work was supported by the Engineering and Physical Sciences Research Council (EPSRC) Centre for Doctoral Training in Fluid Dynamics at the University of Leeds under Grant No. EP/L01615X/1.

\clearpage

\bibliographystyle{apalike}
\bibliography{Master_ref}
\end{document}